%% file: 00head.tex
\documentclass[12pt,a4paper]{article}
\usepackage[left=2.5cm,right=2.5cm,top=1.5cm,bottom=1.5cm]{geometry}
\usepackage{amsmath,comment,cases}
\usepackage{graphicx}
\usepackage{appendix}
\providecommand*{\eu}{\ensuremath{\mathrm{e }}} 
\providecommand*{\iu}{\ensuremath{\mathrm{i }}} 
\usepackage{tikz,pgfplots}  
\usepackage{setspace}
\usepackage{breakcites}
\input{00defns}

\newlength{\figh}
\newlength{\figw}
\begin{document}
\begin{center}
{\Large\bf 
Amplitude decay of Solitary Waves}

{\large P.W. Hammerton \& D.K. Grundy}
\end{center}

\input{sec1}

\input{sec2}

\input{sec3a}

\input{sec3b}

\input{sec3c}

\input{sec3d}

\input{sec4a}

\input{sec4b}

\input{sec4c1}

\input{sec4c2}

\input{sec5}

\appendix
\appendixpage

\input{app_constraint}

\input{app_int}

\input{app_step}

\bibliography{00biblioBKDV}

\end{document}

%% file: 00defns.tex
\newcommand{\sech}{\>\mbox{sech}}

\newcommand{\thalf}{{\textstyle{\frac{1}{2}}}}

\renewcommand{\tt}{{\tilde t}}
\newcommand{\JH}{{\widehat{J}}}
\newcommand{\JB}{{\widetilde{J}}}

\newcommand{\intinf}{\int_{-\infty}^{\infty}}
\newcommand{\Ai}{{\mbox{Ai}}}

\newcommand{\heps}{\delta}

\newcommand{\muno}{{\textstyle\frac{8}{15}}}
\newcommand{\bbJ}{{\overline J}}

\newcommand{\uu}{{\eta^*}}
\newcommand{\uv}{\eta}
\newcommand{\uvh}{\widehat{\eta}}
\newcommand{\uw}{f}

\newcommand{\dx}{{x^*}}
\newcommand{\dt}{{t^*}}
\newcommand{\G}{\gamma}
\newcommand{\cc}{{\widetilde c}}

%% file: sec1.tex
\section{Introduction}

Modelling of the surface elevation of long-wavelength, small-amplitude disturbances on a fluid layer leads to the Korteweg-de Vries (KdV) equation \cite{kdv1895}
\begin{equation}
\uw_t + 6\uw\uw_x + \lambda\uw_{xxx}=0.
\label{eqn:kdv0}
\end{equation}
Including uniform surface tension modifies $\lambda$, the coefficient of dispersion. However, if the Bond number, which characterises the
relative importance of surface tension forces, is close to a critical value, an additional fifth-order spatial derivative term arises \cite{hunter83}. More recently a governing equation was derived taking account of viscous dissipation due to the base boundary condition and arbitrary stress conditions applied at the free surface \cite{hammerton13b}.  In each case the governing equation can be written as a perturbed KdV equation of the form
\[
\uw_t + 6\uw\uw_x + \lambda\uw_{xxx}=\epsilon{\cal R}(\uw),
\]
where additional physical processes described above are governed by the
${\cal R}(\uw)$ term.

The unperturbed KdV equation is well known for having soliton solutions. The single soliton solution of (\ref{eqn:kdv0}) takes the form
\begin{equation}
\uw(x,t)=A\sech^2\left(\alpha(x-ct)\right), \qquad \text{with}\quad \alpha=\sqrt{\frac{A}{2\lambda}},
\label{eqn:soliton0}
\end{equation}
where the propagation speed is given by $c=2A$. In \cite{hammerton13b} numerical solutions of the perturbed KdV equation were presented taking the soliton solution as an initial condition. A wide range of behaviour is seen, which can not be immediately explained in physical terms. For this reason, investigation of the case when the perturbation is small (ie $\epsilon\ll 1$) seems to be a good starting point.

Solutions of the equation
\begin{equation}
\uw_t + \alpha_1\uw\uw_x + \alpha_2\uw_{xxx}=
\epsilon{\cal R}(\uw),
\label{eqn:kdvpert}
\end{equation}
are considered by \cite{ott69,ott70} for $\epsilon\ll 1$. Four different linear terms ${\cal R}(\uw)$ are considered covering the physical situations of magnetosonic waves damped by electron-ion collisions, ion sound waves damped by ion-neutral collisions and by electron Landau damping, and shallow water waves damped by viscosity. The final case previously having been considered by Keulegan \cite{keulegan48}, though without casting the physical problem in the form of a perturbed KdV equation. The approach taken is to consider the solution in the form (\ref{eqn:soliton0}) but with $A$ a slowly varying function of time, with its dependency on $t$ determined by a solvability condition. This is often known as the adiabatic approximation and is explained in more detail in \S\ref{sec:pert}. The advantage of this approach is that the leading-order variation of $A$ with $t$ is easily determined, however changes in the structure of the waveform other than amplitude decay are not identified. Subsequent papers using multi-scale perturbation analysis considered higher order effects, \cite{johnson73,ko78,grimshaw79,grimshaw93,grimshaw03}. These approaches mean that the change in the shape of the waveform can be calculated along with higher order corrections to the wave amplitude and propagation speed.

An alternative approach is to combine perturbation theory and inverse scattering transforms. When $\epsilon=0$ the inverse scattering method \cite{drazin89} provides an exact solution of (\ref{eqn:kdvpert}) for arbitrary initial conditions. In a sequence of papers (\cite{karpman77a,karpman78, karpman79} the first order perturbation in $\epsilon$ to the soliton solution was obtained in the form of an integral for arbitrary 
${\cal R}(\uw)$. Similar methods were also applied to other perturbations of the KdV equation \cite{kaup78}. In \cite{knickerbocker80} explicit results for the case 
${\cal R}(\uw)=-\Gamma(t)\uw$ are obtained and compared with numerical solutions. In all cases the small perturbation was found to have three major effects: (a) a slow change in wave parameters; (b) a deformation of the solitary wave shape and (c) the formation of a small amplitude tail behind the soliton core consisting of a constant `shelf' followed by a region of oscillatory decay. The advantage of inverse scattering method is that in theory more information about the evolution of the disturbance is possible. The disadvantage is that for a given perturbation $\epsilon{\cal R}(\uw)$, obtaining the explicit form of the solution for $\uw(x,t)$ is very cumbersome. With extensive use of algebraic computer packages, the results from inverse scattering theory can be used to find an explicit form for the the core solution, but obtaining expressions for the soliton tail is still difficult. In addition it is unclear whether non-uniformities arise in the perturbation solution which restrict its validity.

The aim of the present paper is to draw together the earlier analyses of perturbed KdV equations, illustrating how the two approaches of 
multi-scale perturbation theory and inverse scattering theory complement each other. We choose to focus on the Burgers-Korteweg-de Vries (BKdV) equation when the perturbation is of the form ${\cal R}(f)=f_{xx}$  in order to provide explicit results. The BKdV equation is widely used the field of wave propagation through cosmic plasmas, see for example the introduction to \cite{gao15}. Particular examples include the propagation of ion-acoustic and magnetosonic waves, with $f$ denoting 
perturbations in either ion  velocity,  ion density or electrostatic wave potential depending on the exact context.
Another application of the BKdV equation is  propagation of gas `slugs' through fluidized beds \cite{harris94} when $f$ describes the voidage fraction, though in this case the perturbation term has the opposite sign and no longer represents a dissipation term. However, in the present analysis we treat  BKdV  as a model equation without considering the significance of the solutions to the physical processes discussed above.

The layout is as follows. In \S 2, the problem is formulated and expressed in a form amenable to asymptotic analysis. In \S 3 the solution is analysed for two separate stages of wave evolution, focussing on the structure of the solution and the wave amplitude. Numerical solutions are discussed in \S 4, and the asymptotic results are validated. Finally in \S 5, comparisons are made with the earlier work outlined in this introduction.

%% file: sec2.tex
\section{Formulation} 

We consider the Burgers-Korteweg-de Vries (BKdV) equation
\[
\eta^*_{\dt}+ 6\eta^*\eta^*_\dx + \lambda\eta^*_{\dx\dx\dx}=
\epsilon^*\eta^*_{\dx\dx},
\]
with boundary conditions $\uu\to 0$ as $\dx\to\pm\infty$. If $\epsilon^*=0$ then travelling wave solutions exist of the form
\[
\uu=2\lambda\alpha_0^2\sech^2(\theta), \qquad
\theta=\alpha_0(\dx-\xi^*), \qquad
\xi^*_\dt=4\lambda\alpha_0^2.
\]
In this paper we consider the  initial condition 
$
\uu(x^*,0)=2\lambda\alpha_0^2\sech^2(\alpha_0 x^*),
$
and solve for $t>0$, corresponding to the model problem of taking a travelling wave solution and switching on the damping term at $t=0$.
Analysis is simplified by non-dimensionalising, setting 
$t^*=t/\lambda\alpha_0^3$, $\uu=\lambda\alpha_0^2\uv$,
$x^*=x/\alpha_0$  and $\epsilon^*=\lambda\alpha_0\epsilon$ to give
\begin{equation}
\uv_t + 6\uv\uv_x + \uv_{xxx}=
\epsilon\uv_{xx}, 
\label{eq:bkdv1}
\end{equation}
with initial condition $\uv(x,0)=2\sech^2(x)$.
We then seek solutions in the case $\epsilon  > 0$ in the form
\[
\uv=2\G^2W(\theta,t),
\qquad
\theta=\G(x-\xi-\chi)
\qquad
\xi_t=4\G^2,
\]
where $\G=1$ and $W=\sech^2\theta$ at $t=0$ and $W$ satisfies
\[W_t+
\G^3\left(
W_{\theta\theta\theta}+12WW_\theta-4W_\theta\right)=
\epsilon\G^2 W_{\theta\theta}-\tfrac{2\G_t}{\G}W-
\tfrac{\G_t}{\G}\theta W_\theta+\G\chi_t W_\theta. \]
Recalling that when $\epsilon=0$, the amplitude, wave number and propagation speed are constant.  When $\epsilon\ll 1$, we seek a solution where $\G,\chi$ are functions of a slow time $\tau=\epsilon t$, in which case
\[
W_t+
\G^3\left(
W_{\theta\theta\theta}+12WW_\theta-4W_\theta\right)=
\epsilon\G^2 \left(
W_{\theta\theta}+\mu\left\{ W+(\theta W)_\theta\right\}+\mu_1W_\theta\right),
\]
where 
\refstepcounter{equation}\label{eq:mudefn}
\begin{equation}
\mu=-\frac{\G_\tau}{\G^3},
\qquad 
\mu_1=\frac{\chi_\tau}{\G}.
\tag{\theequation a,b}
\end{equation}
At this stage $\mu$ and $\mu_1$ are functions of $\tau$ to be determined.

We are interested in the perturbation away from the $\epsilon=0$ solution and so, writing the solution in the form
\begin{equation}
W=F(\theta)+\heps H(\theta,\tt), \qquad F=\sech^2(\theta), \qquad
\heps=\frac{\epsilon}{\G}, \qquad
\tt=\int_0^t \G^3 \> \textrm{d}t,
\label{eq:pert_exp}
\end{equation}
and noting that $\heps_\tt=\mu\heps^2$, gives
\[
H_\tt=\left\{R(F)-L(H)\right\}+\heps
\left\{R(H)-\mu H-6(H^2)_\theta\right\},
\]
where 
\begin{equation}
L(V)=V_{\theta\theta\theta}+
12(FV)_\theta-4V_\theta,
\qquad
R(V)=V_{\theta\theta}+\mu(V+(\theta V)_\theta)+\mu_1V_\theta.
\label{eq:LRdefn}
\end{equation}
The reason for the non-standard definition of the small parameter $\heps$ as a function of $\tau$, and for the scaling of the new time variable $\tt$, is so that the full perturbation equation involves only the one parameter, $\delta$.
Looking now at the small $\heps$ expansion with
$H=J(\theta,\tt)+\heps K(\theta,\tt)+O(\heps^2)$, the first two perturbation terms satisfy
\refstepcounter{equation}\label{eq:jpde}
\begin{equation}
J_\tt=
-L(J)+R(F), \qquad
K_\tt=
-L(K)+R(J)-\mu J-6(J^2)_\theta,
\tag{\theequation a,b}
\end{equation}
with $J(\theta,0)=K(\theta,0)=0$.

When comparing the predictions of asymptotic analysis with numerical results in \S\ref{sec:num}, one key comparison is the maximum amplitude  of the solution,
$\uv_\textsc{m}$, and its position, $x_\textsc{m}$, as  functions of time. 
The maximum is located at 
$\theta_\textsc{m}=\thalf\heps J_\theta(0,\tt)+O(\heps^2)$, 
and hence correct to the order $\epsilon$ term,
\begin{equation}
\uv_\textsc{m}(t)=2\G^2
\left(1+\epsilon \frac{J(0,\tt)}{\G}\right), \qquad
x_\textsc{m}=\frac{1}{\epsilon} \int_0^{\epsilon t}
(4\G^2+\epsilon\G\mu_1)\textrm{d}\tau+
\frac{J_\theta(0,\tt)}{2\G^2}\epsilon.
\label{eq:umax1}
\end{equation}
In the next section we analyse the evolution of  $J$ with time, focussing in particular on the validity of the perturbation expansions over different timescales.

%% file: sec3a.tex
\section{Analytic Solutions of the Linear Perturbation Equation}
\label{sec:pert}

In this section we consider solutions of the perturbation equation
\begin{equation}
J_\tt=
-L(J)+R(F), \qquad
J(\theta,0)=0,
\label{eq:JfinalODE}
\end{equation}
for different ranges of time.
Here the operators $R(.), L(.)$ are defined in (\ref{eq:LRdefn}) and the scaled time variable $\tt$ is defined in terms of $t$ in (\ref{eq:pert_exp}).

%% file: sec3b.tex
\subsection{Solution for $t=O(1)$}
\label{ssec:pert2}

Before considering the solution for $J$ when $t=O(1)$ we 
note that when $J$ satisfies (\ref{eq:JfinalODE}) then there are three integral constraints on $J$, 
\refstepcounter{equation}\label{eq:ivals}
\begin{equation}
\int_{-\infty}^\infty  F J\> \textrm{d}\theta=
2\left(\mu-\muno \right)\tt,
\qquad
\int_{-\infty}^\infty  J\> \textrm{d}\theta=2\mu\tt, \qquad
\int_{-\infty}^\infty  \theta J\> \textrm{d}\theta=
8(\mu-{\textstyle{\frac{4}{5}}})\tt^2-2\mu_1\tt.
\tag{\theequation a,b,c}
\end{equation}
These constraints are derived in Appendix \ref{app:iconstraint}.
One thing that can be immediately concluded from these relations is that there is no time-independent solution which decays as $\theta\to\pm\infty$. However, it is still worthwhile to consider what stationary solutions are possible which are bounded in space, and how they must be modified to simultaneously satisfy the three integral constraints.

\subsubsection{Stationary solution}
\label{ssec:pert_core}

We begin by considering the solution 
$J=\JH(\theta)$
 in which case $\JH_\tt=0$ and (\ref{eq:jpde}a) becomes
 \begin{equation}
\JH_{\theta\theta\theta}+
12(F\JH)_\theta-4\JH_\theta=
F_{\theta\theta}+\mu(F+(\theta F)_\theta)+\mu_1F_\theta.
\label{eq:jode}
\end{equation}
Integrating once with respect to $\theta$ it can be noted that 
$F_\theta=-2\tanh\theta\sech^2\theta$ is a homogeneous solution of the second order linear equation obtained. By writing $\JH=G(\theta)F_\theta$, and solving the second order equation for $G_\theta$  the general solution for $\JH(\theta)$ is finally obtained,
\begin{eqnarray*}
\JH&=&
a\cosh^2\theta+
\left(\frac{\mu}{8}-\frac{1}{15}\right)\tanh\theta\cosh^2\theta+b\tanh\theta+
\left(b-\frac{\mu}{8}+\frac{1}{5}\right)(1-\tanh\theta)
\\
&&\quad
+\,\frac{\mu}{4}\theta\sech^2\theta-\frac{\mu}{8}\theta^2\sech^2\theta~\tanh\theta+
d(\theta\tanh\theta-1)\sech^2\theta+c\tanh\theta\sech^2\theta.
\end{eqnarray*}
Here $a,b,c$ are arbitrary constants and 
$d=3b-\frac{3}{8}\mu-\frac{1}{4}\mu_1+\frac{3}{5}$.
To ensure that $\JH$ does not grow exponentially 
as $\theta\to\pm\infty$ the coefficients of the first two terms must be set to zero and hence 
$a=0$ and $\mu=\muno$.

Looking at the next two terms it is clear that the boundary condition $\JH\to 0$ can not be satisfied at both $\theta=\pm\infty$. It is to be expected that the disturbance tends to zero rapidly in front of the propagating disturbance and hence we set 
$b=0$. This assumption is validated by the numerical results presented in \S\ref{sec:num}. With these conditions imposed,  the stationary solution takes the form
\begin{eqnarray}
\JH &=&\tfrac{1}{15}\left(
2(1-\tanh\theta)+\sech^2\theta\left(2\theta+
(6-{\textstyle \frac{15}{4}}\mu_1)(\theta\tanh\theta-1) -
\theta^2\tanh\theta\right)\right)
 \nonumber
 \\
&&\qquad +
c\tanh\theta\sech^2\theta.
\label{eq:jh0}
\end{eqnarray}
Since $\mu=\frac{8}{15}$, from (\ref{eq:ivals}a) 
$\intinf \sech^2\theta J \textrm{d}\theta=0$. Imposing this condition on $\JH$, 
since $\sech^2\theta$ is exponentially small away from the core region,
and using the standard integral identities in Appendix \ref{app:integrals}  fixes
$\mu_1=\frac{8}{15}$.
Thus $\JH$ is determined apart from the coefficient 
$c$,
\begin{eqnarray}
\JH &=&
\JH_0+c\tanh\theta\sech^2\theta,
\nonumber
\\
\JH_0 &=&
\tfrac{1}{15}\left(
2(1-\tanh\theta)+\sech^2\theta\left(2\theta+
4(\theta\tanh\theta-1)-
\theta^2\tanh\theta\right)\right).
\label{eq:jh}
\end{eqnarray}
Moreover, since $\mu=\mu_1=\muno$, $\G(\tau)$ which describes the evolution of the wave amplitude and wave number is given by solving (\ref{eq:mudefn}a), 
\begin{equation}
\frac{\textrm{d}\G}{\textrm{d}\tau}=-\frac{8\G^3}{15} \quad\Longrightarrow\quad
\G=\frac{1}{(1+\frac{16}{15}\tau)^{\frac{1}{2}}} ,
\label{eq:Gdefn}
\end{equation}
with the speed of propagation given by 
$\xi_t+\epsilon\chi_\tau=4\G^2+\muno\G\epsilon$. The maximum disturbance amplitude is then given by
\[
\uv_\textsc{m}=\frac{2}{1+\frac{16}{15}\tau}\left(1+\epsilon\JH(0,\tt)+O(\epsilon^2)\right)
\sim
2\left(1+\epsilon\left(\JH(0,\tt)-{\textstyle{\frac{16}{15}}}t\right)\right),
\quad\mbox{as ~~$\epsilon t\to 0$}.
\]

However, as previously noted, this solution does not satisfy the required boundary as $\theta\to -\infty$, indeed as 
$\theta\to -\infty$, $\JH\to \frac{4}{15}$, and hence we write
\[
J=\left\{
\begin{array}{ccc}
\JB(\theta,\tt), &\qquad& \theta<-\theta_\textsc{m},
\\
\JH(\theta),&\qquad& \theta > -\theta_\textsc{m}.
\end{array}
\right.
\]
Here $\JB(\theta,\tt)$ describes the transition from 
$0$ as $\theta\to -\infty$ to $\frac{4}{15}$ at
$\theta=-\theta_\textsc{m}$, which we term the tail region. 
The value of $\theta_\textsc{m}$ will be discussed in due course, but 
in \S\ref{sec:num}, numerical examination of $\JH(\theta)$ as $\theta\to -\infty$ reveals it is  within 1\% of its limiting value for $\theta < −4$.

Matching of the stationary (or core) solution to the tail solution is described in \S\ref{ssec:pert_match}, and involves using the integral constraints 
(\ref{eq:ivals} b,c). The contribution to these integrals from the core solution is readily evaluated using the identities in Appendix \ref{app:iconstraint},
\refstepcounter{equation}\label{eq:jhlim}
\begin{equation}
\int_{-\theta_\textsc{m}}^\infty  \JH\> \textrm{d}\theta
\sim 
\tfrac{4}{15}(\theta_\textsc{m}-1)
\qquad
\int_{-\theta_\textsc{m}}^\infty  \theta\JH\> \textrm{d}\theta
\sim 
\tfrac{1}{15}
\left(
\tfrac{\pi^2}{4}-2\theta_\textsc{m}^2\right)+c,
\quad\mbox{as $\theta_\textsc{m}\to\infty$.}
\tag{\theequation a,b}
\end{equation}

\subsubsection{Tail Solution}
\label{ssec:pert_tail}

In the tail region, $F(\theta)$ is exponentially small, so we consider the solution of
\[
\JB_\tt+
\left(
\JB_{\theta\theta\theta}-4\JB_\theta\right)
=0,
\]
with $\JB\to 0$ as $\theta\to -\infty$ and 
$\JB\to \frac{4}{15}$, $\JB_\theta \to 0$ as $\theta\to -\theta_\textsc{m}$
in order to match to the core stationary solution.

First writing $y=\theta+4\tt+D$, with $D$  an as yet undetermined constant, the tail solution satisfies $\JB_\tt+\JB_{yyy}=0$. 
Then in terms of a similarity variable, $z$, $\JB$ satisfies
\begin{equation}
T\JB_{T}+(\JB_{zzz}-z\JB_z)=0,
\quad\mbox{where}\quad 
z=\frac{\theta+4\tt+C}{T}, 
\quad\mbox{and}\quad 
T=(3\tt)^{\frac{1}{3}}.
\label{eq:jb}
\end{equation}
Thus as $\tt$ increases, $z$ increases for fixed $\theta$ and 
the matching condition for $\JB$ becomes
$\JB\to\frac{4}{15}$ and $\JB_z\to 0$ as $z\to\infty$.
The solution of (\ref{eq:jb}) can be obtained by
taking the Fourier transform with respect to $z$. However, a more concise derivation is possible by observing that $\JB_z=\Ai(z)$ is one solution and
writing $\JB$ as a spatial convolution with the Airy function, $\JB=g*\Ai$. The function $g(z,T)$ is obtained by 
substituting into (\ref{eq:jb}), 
to give $Tg_T=zg_z$. This is satisfied if $g(z,T)=\psi(Tz)$ 
for any $\psi$ such that the convolution integral exists. Thus  the leading order solution for the soliton tail is given by
\[
\JB(z,\tt)=\int_{-\infty}^\infty 
\psi\left(T(z-y)\right) \Ai(y)\> \textrm{d}y.
\]
When comparing the analytic solution in the tail region with numerical solutions, it is more convenient to consider the first derivative of $J$ since the matching condition is $\JB_z\to 0$ as $z\to\pm\infty$. Using the solution above we can then write
\begin{equation}
\JB_z(z)=\int_{-\infty}^\infty 
T\phi\left(T(z-y)\right) \Ai(y)\> \textrm{d}y,
\label{eq:jbz}
\end{equation}
where 
$\psi'(z)=\phi(z)$.

In the next section we consider matching of the tail solution to the core solution, and so the asymptotic forms of $\JB$, $\int^z \JB \textrm{d}z$ and 
$\int^z z \JB \textrm{d}z$ for $z\gg 1$ are required. 

\subsubsection{Matching}
\label{ssec:pert_match}

Summarising the structure of the solution constructed,
\begin{equation}
J(\theta,T)=\left\{
\begin{array}{lcc}
\JB =\int_{-\infty}^\infty \psi(Ty) \Ai(z-y)\> \textrm{d}y,
&\qquad& \theta<-\theta_\textsc{m},
\\[12pt]
\JH=\JH_0
+c\tanh\theta\sech^2\theta,
&\qquad& \theta>-\theta_\textsc{m},
\end{array}
\right.
\label{eq:jresult}
\end{equation}
where 
\[\psi(x)=\int_{-\infty}^x \phi(x')\> \textrm{d} x', \quad \textrm{and} \quad z=T^{-1}(\theta+4\tt+C)\] and $\JH_0$ is given by (\ref{eq:jh}).
In the region about $\theta=-\theta_\textsc{m}$ both the functions are constant and it is in this region that the matching occurs, which fixes 
$K_0=\frac{4}{15}$. The constants $c$ and $C$ are now determined in terms of $\phi(x)$ using the integral constraints.

From (\ref{eq:jhlim}a) and (\ref{eq:jhlim}b),
\begin{eqnarray*}
\int_{-\infty}^\infty  J\> \textrm{d}\theta &=&
\int_{-\theta_\textsc{m}}^\infty  \JH\> \textrm{d}\theta+
T\!\!\int_{-\infty}^{z_\textsc{m}}  \JB\> \textrm{d}z
=\tfrac{4}{15}(\theta_\textsc{m}-1)+\tfrac{4}{15}(-\theta_\textsc{m}+4\tt+C)-K_1
\\
&=&\tfrac{16}{15}\tt+\tfrac{4}{15}(C-1)-K_1,
\end{eqnarray*}
which is consistent with (\ref{eq:ivals}b) if  $C=1+\frac{15}{4}K_1$. However at this point we observe that a constant horizontal shift in $g(z)$ (replacing $z$ by $z+Z$)
gives the same shift in 
$\JB(z)$ while the change in $K_1$ is $\frac{4}{15}Z$. Hence the tail solution is unaltered by the value of  $K_1$, so we choose  $K_1=0$
and hence
\begin{equation}
z=(3\tt)^{-\frac{1}{3}}(\theta+4\tt+1).
\label{eq:zdefn}
\end{equation}
Similarly, using (\ref{eq:jhlim}b) and (\ref{eq:jhlim}c),
\begin{eqnarray*}
\int_{-\infty}^\infty  \theta J\> \textrm{d}\theta
&=&
-\tfrac{32}{15}\tt^2-\tfrac{16}{15}\tt+c-
\tfrac{1}{15}
\left(2-\tfrac{1}{4}\pi^2\right)-\tfrac{1}{2}K_2.
\end{eqnarray*}
This is consistent with the third integral constraint (\ref{eq:ivals}c) if
$
{\textstyle c=\frac{1}{15}\left(2-\frac{1}{4}\pi^2\right)+\frac{1}{2}K_2}.
$
Thus assuming that the function $\phi(x)$ which determines the tail solution is known, then the perturbation solution governed by (\ref{eq:jpde}) is known for all $\tt$ once the stationary form of the core is reached. However, without knowledge of the small $\tt$ solution, 
$\phi(x)$ is undetermined, except that
\[ 
\intinf \phi(x) \textrm{d}x=\tfrac{4}{15}
\qquad
\intinf x\phi(x) \textrm{d}x=0,
\]
and the core solution is related to $\phi(x)$ via
\begin{equation}
c=\tfrac{1}{15}\left(
2-\tfrac{1}{4}\pi^2\right)+\tfrac{1}{2}\intinf x^2\phi(x) \textrm{d}x.
\label{eq:c}
\end{equation}

In \S\ref{sec:num}  we compare these asymptotic predictions with numerical results, focussing in particular on the development of the tail behind the main disturbance and the maximum amplitude 
$\uv_\textsc{m}$. However first the validity of this composite description must be considered, as $t$ increases.
In the core region, it has been demonstrated that the perturbation 
$\heps J$ is small compared with the leading order term, as long as 
$\heps=\G^{-1}\epsilon$ is small. Since 
$\G\sim (\epsilon t)^{-\frac{1}{2}}$ 
as $t\to\infty$, the first perturbation term remains small compared with the leading order term until $t=O(\epsilon^{-3})$. However it is not guaranteed that the next perturbation term $\heps^2 K$ is smaller that $\heps J$ -- that is, the first non-uniformity in the  expansion may occur when the second and third terms in the perturbation expansion become comparable in size. 
Indeed considering the equation for $K$ there will be no free parameter in the particular integral of the stationary solution. This points to the presence of a term proportional to $t$ in the core solution for $K$, leading to a non-uniformity in the expansion when $\epsilon t=O(1)$. This is seen more precisely by observing that
\[
\frac{\textrm{d}}{\textrm{d}t}\!\left(\intinf FK \textrm{d}\theta\right)=
\intinf F(R(J)-\mu J-6(J^2)_\theta) \textrm{d}\theta,
\]
where the right-hand side is a non-zero constant. This expression indicates how the breakdown in the asymptotic description can be eliminated and this is described in the next sub-section.

%% file: sec3c.tex
\subsection{Solution for $\tau=\epsilon t=O(1)$}
\label{ssec:pert3}

In the previous section $J(\theta,t)$, the first perturbation away from the leading order soliton solution, was determined by assuming that a core solution develops which is independent of $t$. It was seen that this  represents a small perturbation until $t=O(\epsilon^{-3})$, but that non-uniformity in the perturbation series may arise earlier due to the next term $\heps^2 K$
becoming comparable in size to $\heps J$. This can be analysed by recognising that  the core solution is not truly stationary but evolves on the slow timescale $\tau=\epsilon t$. Thus in the core 
$J=J(\theta,\tau)$ and (\ref{eq:jpde}) becomes
\refstepcounter{equation}\label{eq:jpdenew}
\begin{equation}
L(J)=R(F), \qquad
L(K)=R(J)-\mu J-6(J^2)_\theta-\frac{J_\tau}{\G^2}.
\tag{\theequation a,b}
\end{equation}
The solvability conditions are now subtly different from those used in \S\ref{ssec:pert2}. 
Noting from Appendix \ref{app:iconstraint}, that $\int_{-\infty}^\infty F L(V)\> \textrm{d}\theta=0$ for any function $V(\theta)$ that decays sufficently rapidly to zero as $\theta\to\pm\infty$,
the solvability conditions become
\begin{eqnarray*}
\intinf F(F''+\mu(F+(\theta F)_\theta)+\mu_1F_\theta)\, \textrm{d}\theta &=&0,
\\
\intinf F\left(J''+\mu (\theta F)_\theta+\mu_1 J_\theta
-6(J^2)_\theta-\frac{J_\tau}{\G^2} \right) \textrm{d}\theta &=&0.
\end{eqnarray*}
On evaluating the integrals involving hyperbolic functions, the first of these equations fixes $\mu=\muno$ as before, but $\mu_1$ is left undetermined at this order. The core solution for $J$ is given by (\ref{eq:jh0}) which we write in the form 
\begin{equation}
J={\bbJ}+\mu_1(\tau)g+\cc(\tau)\tanh\theta\sech^2\theta,
\label{eq:J2}
\end{equation}
where
\begin{eqnarray*}
{\bbJ}&=&\frac{1}{15}\left(
2(1-\tanh\theta)+2\theta\sech^2\theta-
\theta^2\sech^2\theta~\tanh\theta-
6(1-\theta\tanh\theta)\sech^2\theta\right),
\\
g&=&\frac{1}{4}(1-\theta\tanh\theta)\sech^2\theta.
\end{eqnarray*}
Substituting into the second solvability condition, all the terms involving $\cc$ cancel out,  and so the evolution of $\cc$ with $\tau$ can not be determined at this order. The remaining terms simplify to 
\[
I_1+\mu_1I_2=\frac{1}{\G^{2}}\frac{\textrm{d}\mu_1}{\textrm{d}\tau} I_3,
\]
where
\begin{eqnarray*}
I_1&=&\intinf F\left(\bbJ_{\theta}+\mu\theta\bbJ
-6\bbJ^2\right)_\theta \> \textrm{d}\theta=\tfrac{176}{225}
\\
I_2&=&\intinf F\left(g_{\theta}+\mu\theta g
-12g\bbJ+\bbJ\right)_\theta \> \textrm{d}\theta=-\tfrac{2}{5}
\\
I_3&=& \intinf Fg \> \textrm{d}\theta=\tfrac{1}{4}.
\end{eqnarray*}
Here the integrals are evaluated using standard results for 
hyperbolic functions and checked using the symbolic manipulation software MAPLE. Using these numerical values and the result 
$\G_\tau=-\muno \G^3$ from (\ref{eq:Gdefn}), the ODE for 
$\mu_1(T)$ can be written as
\[
\G\frac{\text{d}\mu_1}{\text{d}\tau} -3\G_\tau\mu_1=-\tfrac{88}{15}\frac{\G_\tau}{\G^4}
\quad\Longrightarrow\quad \mu_1(\tau)=\tfrac{88}{45}+C\G^3.
\]
The core solution (\ref{eq:J2}) breaks down as $\tau\to 0$, but can be matched to the $t=O(1)$ solution given by (\ref{eq:jh}). This then fixes $\mu_1(0)=\muno$ and hence
\[
\mu_1(\tau)=\tfrac{8}{45}
\left(11-8\G^3\right).
\]
However, $\mu$ does not vary with $\tau$ and hence the expression for $\G(\tau)$ previously derived for $t=O(1)$ remains valid when 
$\tau=O(1)$.
Thus the core solution valid for $\epsilon t=O(1)$ is given by
\[
\uv(x,t)=2\G^2
F(\theta)+2{\G}\left( \bbJ(\theta)+
{\textstyle\muno}(11-8\G^3)g(\theta)\right)
{\epsilon}, \qquad 
\G=(1+{\textstyle\frac{16}{15}}\tau)^{-\frac{1}{2}},
\]
where 
\begin{equation}
\theta=
\textstyle
\G \left(
x+(\frac{15}{2}\log\G )\epsilon^{-1}
-
\left(
{{\frac{11}{3}}}\G^{-1}
+
{{\frac{4}{3}}}\G^2-5\right)
\right).
\label{eq:thetadefn}
\end{equation}
The fact that $\cc(\tau)$ is not determined at this order is hardly surprising since the term $\epsilon c\tanh\theta\sech^2\theta$ in the perturbation series for $W$ can be interpreted as an $O(\epsilon)$ correction to the propagation speed and so would be determined at the next order.

%% file: sec3d.tex
\subsection{Summary of Asymptotic Results}
\label{ssec:asym_summary}

\newcommand{\ttt}{\tt}

The asymptotic analysis presented has demonstrated the solution to be a slowly varying core with propagation speed varying on the slow timescale $\tau$, followed by a tail evolving on the faster timescale and consisting of a near horizontal shelf followed by a decaying oscillation. The structure of the tail is described by a convolution integral involving a single function undetermined by the asymptotic analysis. In the next section this function is determined numerically, but the main means of validating the asymptotic theory  is by considering the maximum amplitude of the wave and its position.
From (\ref{eq:umax1}) the asymptotic prediction of the maximum amplitude for the different timescales, is given by
\begin{subnumcases}
{\uv_\textsc{m}=}
   \displaystyle
\frac{2}{1+\beta}\left(1+\epsilon J(0,\tt)\right)
&
$t=O(1)$
\label{eq:maxb}
\\[12pt]
\displaystyle
\frac{2}{1+\beta}\left(1+\epsilon\left({\textstyle{\frac{2}{9}}}
(1+\beta)^{\frac{1}{2}}-
{\textstyle{\frac{16}{45}}}(1+\beta)^{-1}\right)\right)
&$t=O(\epsilon^{-1})$
\label{eq:maxc}
\end{subnumcases}
where 
$\beta=\frac{16}{15}\epsilon t$ as before. 
The corresponding result for the position of the maximum amplitude is given by
\begin{subnumcases}
{x_\textsc{m}=}
\displaystyle
4t+\epsilon\left(
 {\textstyle 
 \frac{1}{2}J_\theta(0,\tt)+\frac{8}{15}t-\frac{32}{15}t^2
 }
 \right)
& 
$t=O(1)$
\label{eq:xmaxb}
\\[12pt]
\textstyle \nonumber
\frac{15}{4}\log(1+\beta)\epsilon^{-1}+
\left({{\frac{11}{3}}}
(1+\beta)^{\frac{1}{2}}+
{{\frac{4}{3}}}(1+\beta)^{-1}-5\right)\\[12pt]
\textstyle
\qquad + \frac{\epsilon}{2} \cc(\tau) (1+\beta)
&
$t=O(\epsilon^{-1})$
\label{eq:xmaxc}
\end{subnumcases}
The functions 
$\JB(0,\tt)$ and $\JB_\theta(0,\tt)$ are obtained from the numerical solution of (\ref{eq:JfinalODE}), with the asymptotic analysis of \S\ref{ssec:pert2} demonstrating  that 
$\JB(0,\tt)\to -\frac{2}{15}$ and $\JB_\theta(0,\tt)\to c$
as $\tt\to\infty$. As previously noted, the function $\cc(\tau)$ can only be determined by considering higher order terms and this is not pursued here, however as $\tau\to 0$, $\cc\to c$ and hence as $\epsilon t\to 0$ (corresponding to $\beta\to 0$) the final results for both maximum amplitude and position match the $t=O(1)$ result.

%% file: sec4a.tex
\section{Numerical Results}
\label{sec:num}

\subsection{Numerical Schemes}
\label{ssec:num_scheme}
\begin{figure}
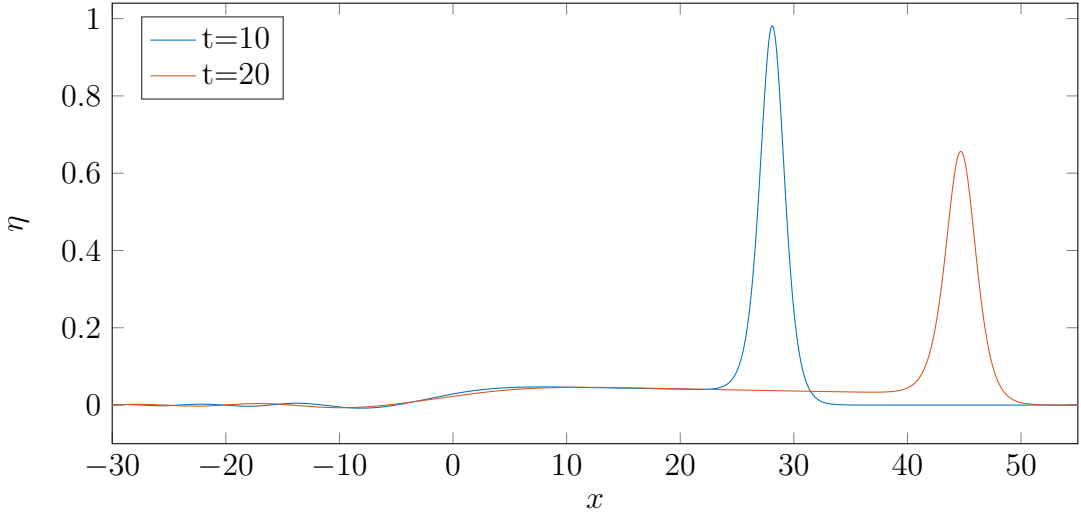

\include{figure1}
\caption{Plot of $\uv(x,t)$ for  $\epsilon=0.1$ at times $t=10,20$, showing the decay in amplitude with increasing $t$ and the development of a decaying tail.}
\label{fig:full}
\end{figure}

Numerical solutions of the BKdV equation (\ref{eq:bkdv1}a) and the linear perturbation equation (\ref{eq:jpde}) are obtained using a pseudospectral scheme \cite{fornberg99} but with linear terms absorbed into an integrating factor following the method of Milewski (\cite{milewski99}). The Fourier transform of (\ref{eq:bkdv1})  takes the form
\begin{equation}
\uvh_t+3\iu k{\cal F}\left[
({\cal F}^{-1}[\uvh])^2
\right]
-\iu k^3\uvh
=-\epsilon k^2\uvh.
\label{eq:ps1}
\end{equation}
Writing
$V=\eu^{h(k)t}\uvh$, where $h(k)=-\lambda\iu k^3+\epsilon k^2$,
(\ref{eq:ps1}) becomes
\[
V_t=-3\iu k \eu^{h(k)t}{\cal F}\left[
\left({\cal F}^{-1}[\eu^{-h(k)t}V]\right)^2
\right].
\]
This scheme was first used to solve the dimensionless BKdV equation (\ref{eq:bkdv1})  with $\epsilon=0.1$, and initial conditions 
$\uv=2\sech^2x$. Guided by the asympotic analysis that predicts a slowly decaying tail behind a  core propagating to the right, a large spatial range $[-400\pi,400\pi]$ was taken, with $N=2^{15}$ spatial points giving a spatial step size $\Delta x\approx 0.04$. Results are presented in Figure \ref{fig:full} for $t=10, 20$, illustrating 
that the main disturbance propagating to the right at  speed $C\approx 4$ with the maximum amplitude decreasing with time. Behind the core is a constant `shelf' extending back to $x\approx 0$, followed by a decaying tail, in agreement with the asymptotics described in \S \ref{sec:pert}. Direct comparison of the numerical results with the analytical results are discussed in \S\ref{ssec:num_bkdv_pert}.
The asymptotic analysis presented in \S\ref{sec:pert} relates to the solution of the perturbation equation (\ref{eq:jpde}a). While this is a linear equation for $J(\theta,\tt)$, a similar scheme to that described above is used. The presence of the $(FJ)_\theta$ term requires two discrete Fourier transforms,
akin to the treatment of the nonlinear term in the first scheme,  with the other terms linear in $J$ being removed using an integrating factor
$\eu^{h_1(k)\tt}$, where $h_1(k)=-\iu k^3-4\iu k$.
The system to be solved is then
\[
V_\tt=-12\iu k \eu^{h_1(k)\tt}{\cal F}\left(
{\cal F}^{-1}(\eu^{-h_1(k)\tt}V)F\right)
+{\cal F}\left( 
F_{\theta\theta}+{\textstyle\frac{8}{15}}(F+(\theta F)_\theta+F_\theta)
\right),
\]
where $F=\sech^2\theta$ as before. Note that the final transform term is independent of $\tt$ and hence is only evaluated once.

%% file: figure1.tex
%
%
\definecolor{mycolor1}{rgb}{0.00000,0.44700,0.74100}%
\definecolor{mycolor2}{rgb}{0.85000,0.32500,0.09800}%
\begin{center}
\begin{tikzpicture}
\begin{axis}[%
width=5.0in,
height=2.30in,
at={(1.091in,0.689in)},
scale only axis,
xmin=-30,
xmax=55,
ymin=-0.1,
ymax=1.04,
axis background/.style={fill=white},
legend style={at={(0.03,0.97)}, anchor=north west, legend cell align=left, align=left, draw=white!15!black},
xlabel={$x$},
ylabel={$\eta$}
]
\addplot [color=mycolor1]
  table[row sep=crcr]{%
-30.0276739228614	0.000880701541568385\\
-29.2223340092215	0.00148415882136987\\
-28.5320426546729	0.00132298656045293\\
-27.7650522607301	0.000447785790761657\\
-26.0393238743588	-0.00179946157842181\\
-25.3873820395074	-0.00181590657732045\\
-24.735440204656	-0.00119480542297623\\
-23.8150517319246	0.000393222943024796\\
-22.7796147001018	0.00205070225754866\\
-22.1276728652504	0.00242758475317828\\
-21.5140805500961	0.00212212588019867\\
-20.8237891955476	0.0010803187426589\\
-18.484468494022	-0.00317599530416146\\
-17.9092256985649	-0.00320301860420358\\
-17.3339829031078	-0.00261815876749694\\
-16.6820410682564	-0.00131919430540961\\
-15.5699049970393	0.00168678329332295\\
-14.7262155637022	0.00368770895236992\\
-14.1126232485479	0.00451992967513348\\
-13.5373804530908	0.00464134620972345\\
-12.9621376576337	0.00408056713214222\\
-12.3485453424794	0.0027949977128614\\
-11.6199044682338	0.000584577888446347\\
-9.51068088489097	-0.00628210741157176\\
-8.85873905003956	-0.00751560216625791\\
-8.24514673488534	-0.00801900762468932\\
-7.6699039394282	-0.00786572524054918\\
-7.09466114397111	-0.0071055714509356\\
-6.48106882881683	-0.00565831245141624\\
-5.82912699396542	-0.00346862320959929\\
-5.10048611971973	-0.000342677203889252\\
-4.29514620607978	0.00376819246999105\\
-3.25970917425701	0.00972964173130464\\
-0.0383495196971353	0.0287543075960954\\
0.920388472731361	0.0335011372799485\\
1.80242742576564	0.0372309678309932\\
2.64611685910273	0.0401797690188133\\
3.52815581213696	0.0426142817712361\\
4.44854428486838	0.0444894237718927\\
5.40728227729687	0.0458028082775712\\
6.48106882881683	0.046632490197112\\
7.74660297882247	0.0469517105880044\\
9.35728280610243	0.0466784890664584\\
11.7349530273252	0.0455681615922359\\
21.7825271879761	0.040277784247202\\
22.3961195031304	0.0407898448286232\\
22.8179642197989	0.0417197826116293\\
23.1631098970732	0.0431207224033656\\
23.4315565349532	0.044840227202485\\
23.661653653136	0.0469522579265487\\
23.8534012516217	0.0493241615132405\\
24.0067993304103	0.0517386475644486\\
24.1601974091989	0.0547264472050131\\
24.3135954879874	0.0584173037384446\\
24.4286440470788	0.0617406314139757\\
24.5436926061703	0.0656263680755771\\
24.6587411652617	0.0701653747446471\\
24.7737897243531	0.0754621418026531\\
24.8888382834446	0.0816364221315382\\
24.9655373228388	0.0863068668522331\\
25.0422363622331	0.0914726975963021\\
25.1189354016274	0.0971827961929819\\
25.1956344410216	0.10349008898455\\
25.272333480416	0.11045169139102\\
25.3490325198102	0.118129005552063\\
25.4257315592045	0.12658775428104\\
25.5024305985988	0.135897931111757\\
25.579129637993	0.146133642364028\\
25.6558286773874	0.157372812949781\\
25.7325277167816	0.169696723192338\\
25.8092267561759	0.183189339407164\\
25.8859257955702	0.197936396656971\\
25.9242753152673	0.205807325421951\\
25.9626248349645	0.214024188322661\\
26.0009743546616	0.222597604843315\\
26.0393238743588	0.231538014441263\\
26.0776733940559	0.240855605958231\\
26.1160229137531	0.250560239831529\\
26.1543724334502	0.260661362784901\\
26.1927219531473	0.271167914716997\\
26.2310714728445	0.282088227555946\\
26.2694209925416	0.293429915911148\\
26.3077705122387	0.305199759430508\\
26.3461200319359	0.317403576864095\\
26.384469551633	0.330046091944318\\
26.4228190713302	0.343130791320341\\
26.4611685910273	0.356659774930399\\
26.4995181107245	0.370633599361483\\
26.5378676304216	0.385051114931478\\
26.5762171501187	0.399909297434085\\
26.6145666698159	0.415203075711112\\
26.652916189513	0.430925156458805\\
26.6912657092101	0.447065847932429\\
26.7296152289073	0.463612884483275\\
26.7679647486044	0.480551254141211\\
26.8063142683016	0.497863031737829\\
26.8830133076959	0.533519604073895\\
26.9597123470901	0.570375221432776\\
27.0364113864844	0.608167224846703\\
27.1514599455758	0.665894037908842\\
27.2665085046673	0.723667315874167\\
27.3432075440616	0.761428825075463\\
27.3815570637587	0.779887878167386\\
27.4199065834558	0.797973353336168\\
27.458256103153	0.815613720152584\\
27.4966056228501	0.832736247889642\\
27.5349551425472	0.849267525046329\\
27.5733046622444	0.865134020881897\\
27.6116541819416	0.880262683412703\\
27.6500037016387	0.894581567151356\\
27.6883532213358	0.908020482761437\\
27.726702741033	0.920511659800667\\
27.7650522607301	0.931990412871393\\
27.8034017804272	0.942395800828137\\
27.8417513001244	0.951671268242514\\
27.8801008198215	0.959765258125493\\
27.9184503395186	0.966631784978262\\
27.9567998592158	0.972230957598775\\
27.995149378913	0.976529441715208\\
28.0334988986101	0.979500853442957\\
28.0718484183072	0.98112607574933\\
28.1101979380044	0.981393491532252\\
28.1485474577015	0.980299128535627\\
28.1868969773986	0.977846713089363\\
28.2252464970958	0.974047631520932\\
28.2635960167929	0.96892079998193\\
28.3019455364901	0.962492445306552\\
28.3402950561872	0.954795801311832\\
28.3786445758844	0.945870726606493\\
28.4169940955815	0.935763251447959\\
28.4553436152787	0.924525062433787\\
28.4936931349758	0.912212934803819\\
28.5320426546729	0.898888122840766\\
28.5703921743701	0.884615719280809\\
28.6087416940672	0.869463994781682\\
28.6470912137643	0.853503728355953\\
28.6854407334615	0.836807539280883\\
28.7237902531587	0.819449230372221\\
28.7621397728558	0.801503151690433\\
28.8004892925529	0.783043592772884\\
28.8388388122501	0.764144210394228\\
28.9155378516443	0.725314299273982\\
28.9922368910386	0.685571020076175\\
29.1839844895243	0.58549797703273\\
29.2606835289186	0.546305969872805\\
29.3373825683129	0.508114745047436\\
29.4140816077071	0.471191519443217\\
29.4524311274043	0.453273380609225\\
29.4907806471015	0.435747070372507\\
29.5291301667986	0.418630755900736\\
29.5674796864957	0.401939620432387\\
29.6058292061929	0.385686032463404\\
29.64417872589	0.369879720988493\\
29.6825282455872	0.354527954212379\\
29.7208777652843	0.339635719432508\\
29.7592272849814	0.325205902078096\\
29.7975768046786	0.311239462166434\\
29.8359263243757	0.297735606701472\\
29.8742758440729	0.284691956788386\\
29.91262536377	0.272104708469328\\
29.9509748834672	0.259968786497758\\
29.9893244031643	0.24827799046119\\
30.0276739228614	0.237025132834511\\
30.0660234425586	0.226202168698229\\
30.1043729622557	0.215800316989274\\
30.1427224819528	0.2058101732663\\
30.18107200165	0.196221814069112\\
30.2194215213472	0.187024893032678\\
30.2577710410443	0.178208728983336\\
30.2961205607414	0.169762386297464\\
30.3344700804386	0.161674747844941\\
30.3728196001357	0.15393458086993\\
30.44951863953	0.139451501057145\\
30.5262176789242	0.126223067282091\\
30.6029167183186	0.114160517827514\\
30.6796157577128	0.103177502699843\\
30.7563147971071	0.0931908978141323\\
30.8330138365014	0.0841213844156385\\
30.9097128758957	0.0758938344996096\\
30.98641191529	0.0684375384332796\\
31.0631109546842	0.0616863063572808\\
31.1398099940785	0.0555784704748419\\
31.2165090334728	0.0500568111706272\\
31.2932080728671	0.0450684261190872\\
31.3699071122614	0.0405645581809182\\
31.4849556713528	0.0346202238318583\\
31.6000042304442	0.0295303360617396\\
31.7150527895356	0.0251762540366087\\
31.8301013486271	0.0214547122667383\\
31.9451499077185	0.01827612923271\\
32.098547986507	0.0147502419776302\\
32.2519460652956	0.011898140452054\\
32.4436936637813	0.00908954157822706\\
32.635441262267	0.0069394796108142\\
32.8655383804499	0.00501590618920034\\
33.1339850183298	0.00343154510817811\\
33.4791306956041	0.0021037327146658\\
33.9009754122727	0.00115488835393762\\
34.5145677274269	0.000481291666133643\\
35.5116552395526	0.000115256962480714\\
38.1577720986553	2.4953375046266e-06\\
55.0315607653974	-3.26849658449646e-13\\
};
\addlegendentry{t=10}

\addplot [color=mycolor2]
  table[row sep=crcr]{%
-30.0276739228614	3.98534233454484e-05\\
-28.3786445758844	0.00179190814020558\\
-27.4966056228501	0.00202437456918148\\
-26.652916189513	0.00159105221271716\\
-25.6558286773874	0.000399792524440556\\
-23.4315565349532	-0.00249138096170043\\
-22.587867101616	-0.0027690389624766\\
-21.7825271879761	-0.00238478742813442\\
-20.9004882349419	-0.00130167452217478\\
-19.5582550455419	0.00110527941829019\\
-18.2927208955363	0.00316682642717581\\
-17.4490314621992	0.00388985074176418\\
-16.6820410682564	0.00389914500915722\\
-15.9150506743135	0.00326160152904009\\
-15.0713612409764	0.00190926195138275\\
-13.9208756500622	-0.000636421452355762\\
-12.0417491849023	-0.00481574814259744\\
-11.1213607121709	-0.00618763612241224\\
-10.2776712788338	-0.0068280046297815\\
-9.4339818454967	-0.00681772578459316\\
-8.59029241215961	-0.00615487696538963\\
-7.70825345912533	-0.00480490780547882\\
-6.78786498639397	-0.00275643786920199\\
-5.75242795457115	0.000207813486930775\\
-4.56359284395978	0.00428587711674311\\
-3.06796157577128	0.0101104552130238\\
2.14757310303992	0.0310362364433132\\
3.45145677274269	0.0352965400002958\\
4.67864140305119	0.0386693604236896\\
5.86747651366255	0.0412946717977434\\
7.05631162427397	0.043282637606552\\
8.28349625458247	0.0447036387956814\\
9.62572944398238	0.0456104121079051\\
11.1213607121709	0.045972698271278\\
12.9237881379365	0.0457503404229982\\
15.3781573985535	0.0447542899619364\\
19.9417502425133	0.04212503052144\\
30.1043729622557	0.0362982644504015\\
36.0485485153126	0.0334591612081212\\
37.3907817047125	0.0333835899299757\\
38.1194225789582	0.0339320421796785\\
38.617966335021	0.0348754158166145\\
39.0014615319924	0.0361564811714175\\
39.3466072092667	0.0379519304011566\\
39.6150538471467	0.0399513398923617\\
39.8451509653295	0.0422337535308017\\
40.0752480835123	0.0451983985684592\\
40.2669956819981	0.0483267300521177\\
40.4203937607866	0.0513568678857865\\
40.5737918395752	0.0549454158710603\\
40.7271899183638	0.0591892710405588\\
40.8805879971524	0.0642004857087031\\
40.9956365562438	0.0685393195834365\\
41.1106851153352	0.073442972132419\\
41.2257336744266	0.0789786226067548\\
41.340782233518	0.0852197745404197\\
41.4558307926094	0.0922463406226157\\
41.5708793517009	0.100144544419486\\
41.6475783910952	0.105939553668009\\
41.7242774304895	0.112191479435268\\
41.8009764698837	0.11892990051556\\
41.877675509278	0.126185138363887\\
41.9543745486723	0.133988030532436\\
42.0310735880666	0.142369646596414\\
42.1077726274609	0.151360938887102\\
42.1844716668551	0.160992320169107\\
42.2611707062494	0.171293160468061\\
42.3378697456437	0.182291195659516\\
42.414568785038	0.194011841267624\\
42.4912678244323	0.206477406303442\\
42.5679668638265	0.219706204014734\\
42.6446659032208	0.233711559242771\\
42.7213649426151	0.24850071580024\\
42.7980639820094	0.264073651995076\\
42.8747630214037	0.280421818189524\\
42.951462060798	0.297526817112349\\
43.0281611001922	0.315359055468178\\
43.1048601395865	0.33387640404564\\
43.1815591789808	0.35302291272275\\
43.2582582183751	0.372727636064745\\
43.3733067774665	0.403136627225578\\
43.5267048562551	0.444682728798071\\
43.7184524547408	0.496625529381539\\
43.795151494135	0.516860390842318\\
43.8718505335293	0.536497435552604\\
43.9485495729236	0.555331223170136\\
44.0252486123179	0.573150922145011\\
44.063598132015	0.581614132952581\\
44.1019476517122	0.589744429476525\\
44.1402971714093	0.597515854232185\\
44.1786466911065	0.604902964136436\\
44.2169962108036	0.611880989891084\\
44.2553457305007	0.618425996458789\\
44.2936952501979	0.624515042998794\\
44.332044769895	0.630126340558959\\
44.3703942895922	0.635239405774307\\
44.4087438092893	0.639835208803994\\
44.4470933289865	0.643896313749607\\
44.4854428486836	0.647407009838908\\
44.5237923683807	0.650353431731574\\
44.5621418880779	0.652723667406413\\
44.600491407775	0.654507852221762\\
44.6388409274722	0.655698247900943\\
44.6771904471693	0.656289305379467\\
44.7155399668665	0.656277710657207\\
44.7538894865636	0.65566241302303\\
44.7922390062607	0.654444635256773\\
44.8305885259579	0.652627865658914\\
44.868938045655	0.650217832007662\\
44.9072875653521	0.647222457789546\\
44.9456370850493	0.643651801290027\\
44.9839866047464	0.639517978357901\\
45.0223361244436	0.634835069868693\\
45.0606856441407	0.629619015102534\\
45.0990351638379	0.623887492418142\\
45.137384683535	0.617659788743516\\
45.1757342032321	0.610956659514002\\
45.2140837229293	0.603800180767571\\
45.2524332426264	0.596213595155739\\
45.2907827623235	0.588221153646067\\
45.3291322820207	0.579847954680226\\
45.3674818017178	0.571119782511147\\
45.4441808411121	0.55270412207912\\
45.5208798805064	0.533188122864722\\
45.5975789199007	0.512786788625228\\
45.674277959295	0.491711735525428\\
45.7893265183864	0.459280042486071\\
46.0577731562664	0.382959088990248\\
46.1728217153578	0.351210279942592\\
46.2495207547521	0.330664874694179\\
46.3262197941464	0.310720671667816\\
46.4029188335406	0.29144902690517\\
46.4796178729349	0.272905947385944\\
46.5563169123292	0.255133287591079\\
46.6330159517235	0.238160062332042\\
46.7097149911178	0.222003818453487\\
46.786414030512	0.206672017440837\\
46.8631130699063	0.192163390342259\\
46.9398121093006	0.178469235284844\\
47.0165111486949	0.165574635913487\\
47.0932101880892	0.153459586128662\\
47.1699092274835	0.142100012467139\\
47.2466082668777	0.131468690361594\\
47.3233073062721	0.121536054389352\\
47.4000063456663	0.112270905572259\\
47.4767053850606	0.103641020933779\\
47.5534044244549	0.0956136719782208\\
47.6301034638491	0.0881560596508919\\
47.7068025032435	0.0812356737796094\\
47.7835015426377	0.0748205850895332\\
47.860200582032	0.068879677711827\\
47.9752491411234	0.0607918209481113\\
48.0902977002149	0.0536065701311159\\
48.2053462593063	0.0472334956620202\\
48.3203948183977	0.0415888756414589\\
48.4354433774892	0.0365957824610916\\
48.5504919365806	0.0321839801150574\\
48.665540495672	0.0282896874663905\\
48.8189385744606	0.0238034999269701\\
48.9723366532491	0.0200145329829482\\
49.1257347320376	0.0168182219660267\\
49.3174823305234	0.0135204770142252\\
49.5092299290091	0.0108613453091664\\
49.7393270471919	0.00834428327156189\\
50.0077736850719	0.00612886718970884\\
50.314569842649	0.00430269253103432\\
50.6597155199233	0.00288643472850936\\
51.119909756289	0.00169211388446655\\
51.7335020714432	0.000827945458574675\\
52.6155410244775	0.000294876547457079\\
54.3029198911517	4.03172951664033e-05\\
55.0315607653974	1.6975727724855e-05\\
};
\addlegendentry{t=20}
\end{axis}
\end{tikzpicture}

\end{center}

%% file: sec4b.tex
\subsection{Numerical results for perturbation equation} 
\label{ssec:num_pert}
\begin{figure}[ht]
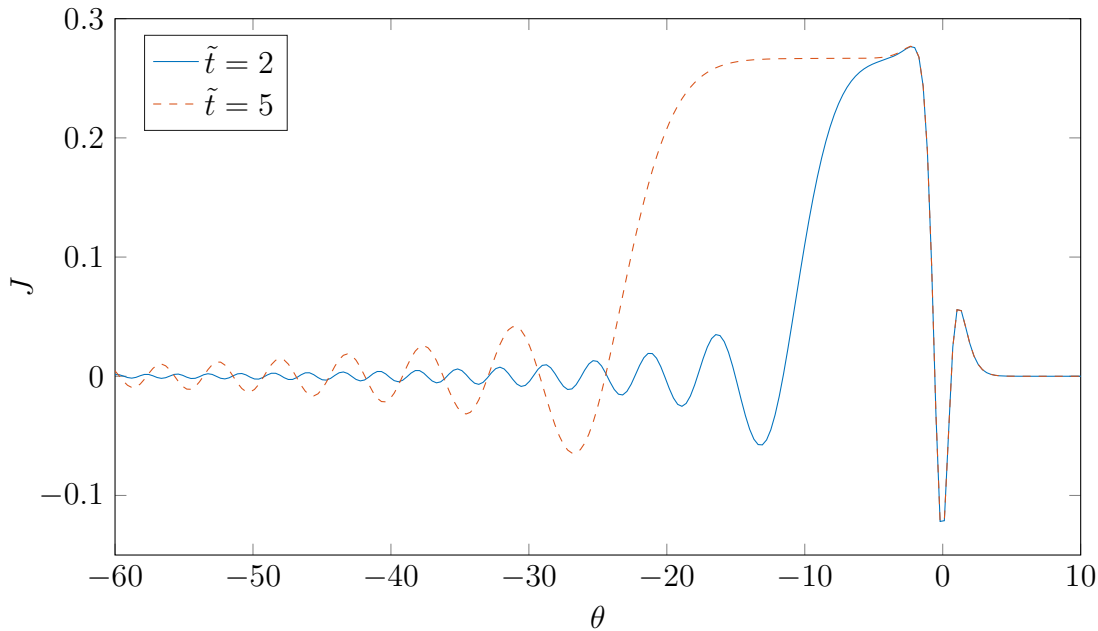

\include{figure2}
\caption{Plot of numerical solution $J(\theta,\tt)$ for $\tt=2$ and $\tt=5$ illustrating the development of a stationary core about 
$\theta=0$, a constant `shelf' behind the core, followed by an oscillating tail. }
\label{fig:j1}
\end{figure}

We now consider 
numerical solutions of the perturbation equation (\ref{eq:jpde}a).
In figure \ref{fig:j1}, numerical results for $J(\theta,\tt)$ are plotted for 
$\tt=2,5$.
The first thing to note is that the core solution (the solution around $\theta=0$) has already approached a stationary form at $\tt=2$.
Ahead of the core, the solution decreases rapidly to zero confirming the analysis presented in  \S\ref{ssec:pert_core}.
Behind the core, a shelf of constant amplitude has appeared by $\tt=5$ and the matching range between core and tail discussed in \S\ref{ssec:pert2} corresponds to the region $-15<\theta<-5$. 
Results for larger $\tt$ show that the extent of the shelf increases as $\tt$ increases.

The stationary form of the solution around $\theta=0$ is now compared with the predicted analytic form of the core (\ref{eq:jh}). 
Focussing first on the value of $J(0,\tt)$, the numerical results confirm the $\tt$ limit $J(0,\tt)\to -\frac{2}{15}$ from (\ref{eq:jh}), with $J(0,\tt)$ attaining 95\% of its limiting value when $\tt=0.7$ and  99\% by $\tt=1.1$.
The undetermined coefficient $c$ appearing in the core solution can be directly extracted from the numerical results in a number of different ways. Directly from the numerical solution, $J_\theta(0,t)\to c$ as $t\to\infty$. Alternatively, from the integral constraint (\ref{eq:jhlim}b)
\[
c \sim 
\int_{-\theta_\textsc{m}}^\infty  \theta\JH\> \textrm{d}\theta
-\frac{1}{15}
\left(
\frac{\pi^2}{4}-2{\theta_\textsc{m}}^2\right),
\]
where $\theta_\textsc{m}$ is taken to be in the overlap region. A third method for calculating $c$,  comes by comparing the computed  value of 
$J-\JH_0$ with $c\tanh\theta\sech^2\theta$ at 
$\theta=\tanh^{-1}(1/\sqrt{3})$, the position of the maximum of $\tanh\theta\sech^2\theta$.
At $\tt=5$ all three methods give $c=0.0451$ correct to 3 significant figures.

\begin{figure}[ht]
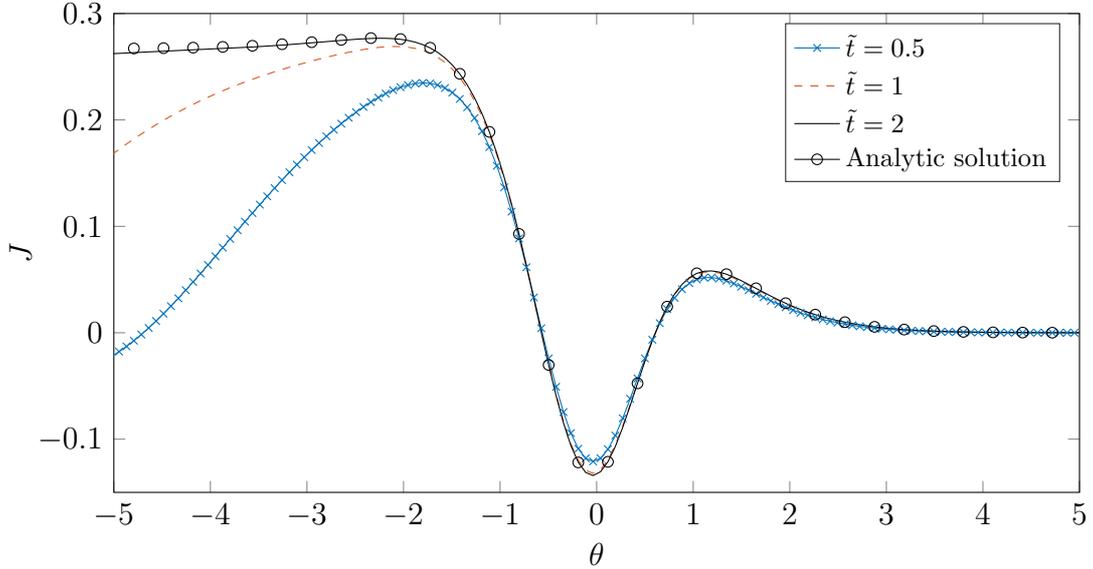

\include{figure3}
\caption{Plot of numerical results for $J(\theta,\tt)$ for $\tt=0.5, 1.0$ and 
$2.0$ along with the analytic solution (\ref{eq:jh})  with $c=0.0451$ (symbols).}
\label{fig:j2}
\end{figure}

In figure \ref{fig:j2} numerical results for $\tt=0.5, 1, 2$ 
are compared  with the analytic results using the computed value of $c$.
It is seen that there is good agreement between numerical and analytic solutions over the main part of the core, even for $\tt=0.5$. When $\tt=2$, results are indistinguishable over the range $-5<\theta<5$.

Looking now at the tail region, it was shown in \S\ref{ssec:pert_tail} that the solution $\JB$
can be written in terms of a single universal function $\phi(x)$,
The method used to extract this function from the numerical solution is to compare the derivative of the numerical solution and that of the analytic solution. From (\ref{eq:jbz}), 
\[
\JB_z=h*\Ai(z), \qquad h(z,\tt)=T\phi(Tz),
\qquad 
T=(3\tt)^{\frac{1}{3}}.
\]
where $z=(3\tt)^{-\frac{1}{3}}(\theta+4\tt+1)$ from (\ref{eq:zdefn}). Taking Fourier transforms with respect to $z$ and using the convolution theorem then gives
\begin{equation}
h(z,\tt)={\cal F}^{-1}\left(\frac{{\cal F}(\JB_z)}{\exp(\iu k^3/3)}\right),
\label{eq:hdefn}
\end{equation}
since ${\cal F}(\Ai)=\exp(\iu k^3/3)$ .
To obtain a numerical approximation of $h$ (and hence $\phi$) we define
$Q(z,t)$  to be  the computed value of $J_z$ for $\theta<\theta_\textsc{m}$ and zero elsewhere, with  $h(z,\tt)$ then given by 
${\cal F}^{-1}\left({\cal F}(Q)/\exp(\iu k^3/3)\right)$
The exact value of the Fourier transform of the Airy function is used rather than the discrete transform over the finite range of the numerical calculation. This proves a better approach since the slow decay of $\Ai(z)$ accompanied by a shortening wavelength as 
$z\to -\infty$, leads to inaccuracy in the large wavenumber components of the discrete  transform of the Airy function over a finite spatial range. 

\begin{figure}[ht]
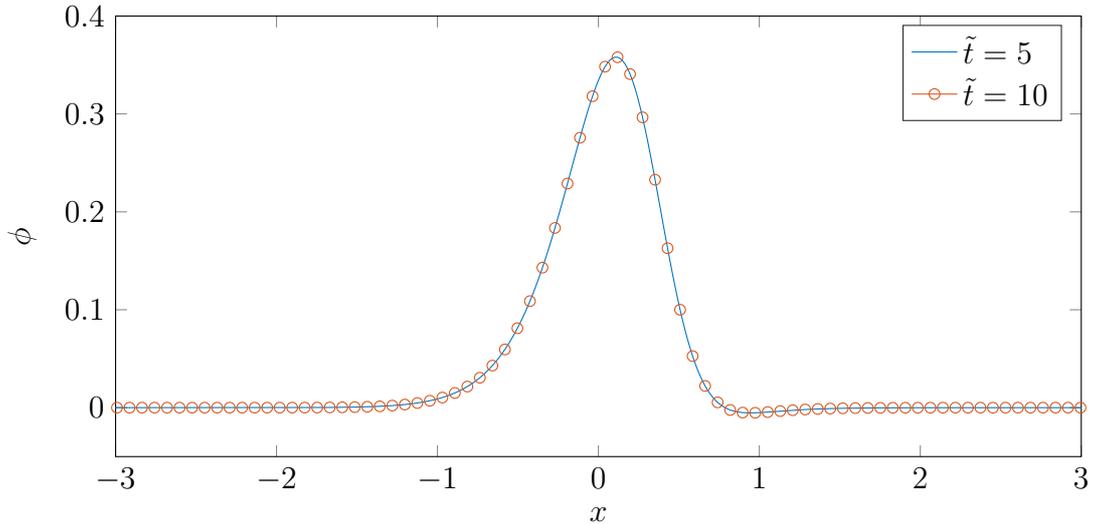

\begin{center}
\include{figure4}
\end{center}
\caption{Plot of $\phi(x)$ for $\tt=5$ (line) and $\tt=10$ (symbols),  where $\phi(x)=h\left(x/(3\tt\right)^{\frac{1}{3}})/(3\tt)^{\frac{1}{3}}$, with $h$  extracted from the numerical solution of $J$ using (\ref{eq:hdefn}).}
\label{fig:phi}
\end{figure}

The function $\phi(x)$ extracted from the numerical solution at $\tt=5$ is shown in figure \ref{fig:phi}. Extraction of $\phi(x)$ at $\tt=10$ produced indistinguishable results. Thus the extracted value of $\phi$ can be used to give the solution in the tail region for all $\tt>5$.

Finally, the computed value of $\intinf x^2\phi(x) \textrm{d}x=0.153$ which using 
(\ref{eq:c}) gives a value of $c=0.0453$, agreeing to within $0.5\%$ of the value extracted from the core solution.

%% file: figure2.tex
%
%
\definecolor{mycolor1}{rgb}{0.00000,0.44700,0.74100}%
\definecolor{mycolor2}{rgb}{0.85000,0.32500,0.09800}%
\begin{center}
\begin{tikzpicture}
\begin{axis}[%
width=5.0in,
height=2.80in,
at={(1.125in,0.63in)},
scale only axis,
xmin=-60,
xmax=10,
ymin=-0.15,
ymax=0.3,
xtick={-60,-50,...,10},
axis background/.style={fill=white},
legend style={at={(0.03,0.97)}, anchor=north west, legend cell align=left, align=left, draw=white!15!black},
xlabel={$\theta$},
ylabel={$J$}
]
\addplot [color=mycolor1]
  table[row sep=crcr]{%
-60.0155473701438	0.00140309285878004\\
-59.7087512125668	0.00131850965405533\\
-59.4019550549895	0.00022108223373607\\
-59.0951588974124	-0.00107220151387821\\
-58.7883627398353	-0.00159422081604532\\
-58.4815665822582	-0.000929987539002752\\
-58.1747704246811	0.000445071044588019\\
-57.8679742671038	0.00153113434727459\\
-57.5611781095267	0.00151399500544613\\
-57.2543819519497	0.000387199560456963\\
-56.9475857943726	-0.00105731209648496\\
-56.6407896367955	-0.00177269581885042\\
-56.3339934792182	-0.00123212405774353\\
-55.720401164064	0.00154213993815944\\
-55.4136050064869	0.00182174017942316\\
-55.1068088489098	0.000823062816685649\\
-54.8000126913325	-0.00077816159888755\\
-54.4932165337555	-0.00188744805239338\\
-54.1864203761784	-0.00172085339040251\\
-53.8796242186013	-0.000369410427119021\\
-53.5728280610242	0.00127708335112686\\
-53.2660319034469	0.00210254726926706\\
-52.9592357458698	0.00153738867078346\\
-52.3456434307157	-0.00168636036622871\\
-52.0388472731386	-0.00223066027744068\\
-51.7320511155613	-0.00132258228069304\\
-51.4252549579842	0.000473307964831804\\
-51.1184588004071	0.00201952646178682\\
-50.81166264283	0.00230895584627433\\
-50.5048664852529	0.00113296838519261\\
-50.1980703276756	-0.000801776900686946\\
-49.8912741700985	-0.00228693421368575\\
-49.5844780125215	-0.00238753146410176\\
-49.2776818549444	-0.00100246597938281\\
-48.9708856973673	0.00104203298232619\\
-48.66408953979	0.00251367165784444\\
-48.3572933822129	0.00249690797208046\\
-48.0504972246358	0.000972378708588906\\
-47.7437010670587	-0.0011868879250585\\
-47.4369049094817	-0.00270382537584624\\
-47.1301087519043	-0.00267199033955734\\
-46.8233125943273	-0.00106697620104512\\
-46.5165164367502	0.00120815001412211\\
-46.2097202791731	0.00286021619709942\\
-45.902924121596	0.00292043413538323\\
-45.5961279640187	0.00132196488831227\\
-45.2893318064416	-0.00107848520610077\\
-44.9825356488645	-0.00294913292977128\\
-44.6757394912875	-0.00324296763124465\\
-44.3689433337104	-0.00175341923445416\\
-44.0621471761331	0.000742165356648172\\
-43.755351018556	0.00292013216905218\\
-43.4485548609789	0.00359436268755076\\
-43.1417587034018	0.00237392765315292\\
-42.5281663882474	-0.0026737829112875\\
-42.2213702306703	-0.00389516686833957\\
-41.9145740730933	-0.00314343655734461\\
-41.6077779155162	-0.000759974400459384\\
-41.3009817579391	0.00209361749002568\\
-40.9941856003618	0.00398900624681175\\
-40.6873894427847	0.00396422722491963\\
-40.3805932852076	0.00197709259063572\\
-40.0737971276305	-0.00104394741045155\\
-39.7670009700532	-0.00366723539855229\\
-39.4602048124761	-0.00461720316456393\\
-39.1534086548991	-0.00340568813862063\\
-38.846612497322	-0.000542189900549772\\
-38.5398163397449	0.00267338792981775\\
-38.2330201821676	0.00477053660500815\\
-37.9262240245905	0.004754950465653\\
-37.6194278670134	0.00258550586453765\\
-37.3126317094363	-0.000826651466041994\\
-37.0058355518593	-0.00399279646497774\\
-36.699039394282	-0.00552165280704031\\
-36.3922432367049	-0.00469969963425854\\
-36.0854470791278	-0.00182555780997973\\
-35.7786509215507	0.00193702617346503\\
-35.4718547639736	0.00502046339192219\\
-35.1650586063963	0.00612861560794897\\
-34.8582624488192	0.0047466613894116\\
-34.5514662912421	0.00137820277659984\\
-34.2446701336651	-0.00267851608035841\\
-33.937873976088	-0.00581973359722099\\
-33.6310778185107	-0.00679064437500188\\
-33.3242816609336	-0.00514928104787771\\
-33.0174855033565	-0.0014631936755265\\
-32.7106893457794	0.00292833271278425\\
-32.4038931882023	0.00638976544210124\\
-32.097097030625	0.00761916252287875\\
-31.790300873048	0.00609978260843036\\
-31.4835047154709	0.00231262848361524\\
-31.1767085578938	-0.00245885967373027\\
-30.8699124003167	-0.00655457746025689\\
-30.5631162427394	-0.00853878947148701\\
-30.2563200851623	-0.00766676700716573\\
-29.9495239275852	-0.00416134329291395\\
-29.3359316124311	0.00585329993898398\\
-29.0291354548538	0.00913123706248342\\
-28.7223392972767	0.00961943836655621\\
-28.4155431396996	0.00709270265281248\\
-28.1087469821225	0.00224441568068556\\
-27.8019508245454	-0.00350221519032345\\
-27.4951546669681	-0.00844930041053971\\
-27.188358509391	-0.0111020811771922\\
-26.881562351814	-0.0106285250702314\\
-26.5747661942369	-0.00706928321626066\\
-26.2679700366598	-0.00134733466248349\\
-25.9611738790825	0.00501782305566678\\
-25.6543777215054	0.0103019361328478\\
-25.3475815639283	0.0130640358328975\\
-25.0407854063512	0.0124976220000192\\
-24.7339892487742	0.00865944025883891\\
-24.4271930911968	0.00242245119118678\\
-24.1203969336198	-0.0047228827887551\\
-23.8136007760427	-0.0110635057759012\\
-23.5068046184656	-0.0150500292219533\\
-23.2000084608885	-0.0156836215626015\\
-22.8932123033112	-0.0127132069058788\\
-22.5864161457341	-0.0067081779978011\\
-22.279619988157	0.00110355836150688\\
-21.97282383058	0.00908361678988712\\
-21.6660276730029	0.0155599548592136\\
-21.3592315154256	0.0191456497303761\\
-21.0524353578485	0.0190384250885955\\
-20.7456392002714	0.0151415790751486\\
-20.4388430426943	0.00809460028941089\\
-20.1320468851172	-0.000887065406189436\\
-19.8252507275399	-0.010219685174107\\
-19.5184545699628	-0.0182658338590471\\
-19.2116584123858	-0.023591757100526\\
-18.9048622548087	-0.0252278375541053\\
-18.5980660972316	-0.0227875052972664\\
-18.2912699396543	-0.0165369714929398\\
-17.9844737820772	-0.00730832733972164\\
-17.6776776245001	0.00361652620690478\\
-17.370881466923	0.0147312510985884\\
-17.064085309346	0.0244961969627653\\
-16.7572891517686	0.0315714566328111\\
-16.4504929941916	0.0349552832720761\\
-16.1436968366145	0.0341190867020202\\
-15.8369006790374	0.029021960953294\\
-15.5301045214603	0.0201162306869733\\
-15.223308363883	0.00824491172401309\\
-14.9165122063059	-0.00544404550846878\\
-14.6097160487288	-0.0196580592137323\\
-14.3029198911518	-0.0330678308157033\\
-13.9961237335747	-0.0444570271202807\\
-13.6893275759974	-0.052793247112767\\
-13.3825314184203	-0.0573229313253591\\
-13.0757352608432	-0.057578818512269\\
-12.7689391032661	-0.0534109976517101\\
-12.462142945689	-0.0449369839915192\\
-12.1553467881117	-0.0325266149296937\\
-11.8485506305346	-0.0167197558834218\\
-11.5417544729576	0.0018068353504006\\
-11.2349583153805	0.022326577182632\\
-10.9281621578034	0.0440899250111713\\
-10.314569842649	0.0886047048049576\\
-10.0077736850719	0.110177602781469\\
-9.70097752749484	0.130673050224893\\
-9.39418136991776	0.149772030924936\\
-9.08738521234045	0.167247658248911\\
-8.78058905476337	0.182982878124228\\
-8.47379289718629	0.196929849876462\\
-8.1669967396092	0.209122236645349\\
-7.86020058203189	0.219632318839203\\
-7.55340442445481	0.228583657303332\\
-7.24660826687773	0.236110407776053\\
-6.93981210930065	0.242373553440402\\
-6.63301595172356	0.247524396842245\\
-6.32621979414625	0.25172539769369\\
-6.01942363656917	0.255118083382051\\
-5.71262747899209	0.257848312373163\\
-5.40583132141501	0.260038091779045\\
-5.09903516383793	0.261814794126209\\
-4.79223900626062	0.263286301761703\\
-3.56505437595229	0.268527949255393\\
-3.25825821837498	0.27031043128796\\
-2.95146206079789	0.272455280473743\\
-2.64466590322081	0.274796796665989\\
-2.33786974564373	0.276577943442824\\
-2.03107358806665	0.275757153959759\\
-1.72427743048934	0.26765883522922\\
-1.41748127291226	0.24323692509293\\
-1.11068511533517	0.188692729061906\\
-0.80388895775809	0.0928703033915284\\
-0.497092800181008	-0.0302140513679703\\
-0.190296642603698	-0.12179040634053\\
0.116499514973384	-0.121202364283569\\
0.423295672550466	-0.0475046109441948\\
0.730091830127549	0.0245468130510602\\
1.03688798770463	0.0558379034526268\\
1.34368414528194	0.05502952916558\\
1.65048030285902	0.041735810951856\\
1.95727646043611	0.0277306494626686\\
2.26407261801319	0.0170340627925327\\
2.57086877559027	0.00992661157548014\\
2.87766493316758	0.00556407555814076\\
3.18446109074466	0.00301555840727019\\
3.49125724832174	0.00158603655776801\\
3.79805340589883	0.000805053020847879\\
4.10484956347591	0.000394850196727248\\
4.7184418786303	7.93600681845419e-05\\
5.94562650893886	-2.89447992685155e-06\\
10.2407727150185	8.60013535941562e-07\\
};
\addlegendentry{$\tilde{t}=2$}

\addplot [color=mycolor2, dashed]
  table[row sep=crcr]{%
-60.0155473701438	0.0047169359095065\\
-59.7087512125668	0.000460762937521508\\
-59.4019550549895	-0.00399725457523203\\
-59.0951588974124	-0.0075696869076296\\
-58.7883627398353	-0.00940883736161879\\
-58.4815665822582	-0.00904268319190749\\
-58.1747704246811	-0.00654832114343407\\
-57.8679742671038	-0.00246719052483968\\
-57.5611781095267	0.00225786684455898\\
-57.2543819519497	0.00656567681633646\\
-56.9475857943726	0.00945396170592971\\
-56.6407896367955	0.0102704751004978\\
-56.3339934792182	0.00879197715263302\\
-56.0271973216411	0.0053368932027027\\
-55.720401164064	0.000626869614258396\\
-55.4136050064869	-0.00430762251826877\\
-55.1068088489098	-0.0084173621322563\\
-54.8000126913325	-0.0107944013803376\\
-54.4932165337555	-0.0109293355027162\\
-54.1864203761784	-0.00875136688208045\\
-53.8796242186013	-0.00470502614630419\\
-53.2660319034469	0.00553947168369717\\
-52.9592357458698	0.0096576317546706\\
-52.6524395882927	0.0119065695454381\\
-52.3456434307157	0.0118319001495166\\
-52.0388472731386	0.0094048979588095\\
-51.7320511155613	0.0050895445846848\\
-51.1184588004071	-0.00576994574821299\\
-50.81166264283	-0.0102525872825296\\
-50.5048664852529	-0.0128920341589662\\
-50.1980703276756	-0.0131913352372521\\
-49.8912741700985	-0.0110510297849089\\
-49.5844780125215	-0.00684975539353161\\
-49.2776818549444	-0.00130558262356573\\
-48.9708856973673	0.00458947851838332\\
-48.66408953979	0.00981338316569946\\
-48.3572933822129	0.0134273882118521\\
-48.0504972246358	0.0148032993871325\\
-47.7437010670587	0.0136624600884971\\
-47.4369049094817	0.0101845440103929\\
-47.1301087519043	0.00490759510777394\\
-46.5165164367502	-0.00744991416292606\\
-46.2097202791731	-0.0125305221751475\\
-45.902924121596	-0.0157333976320402\\
-45.5961279640187	-0.0165137628533145\\
-45.2893318064416	-0.0147373469022156\\
-44.9825356488645	-0.0106306803062211\\
-44.6757394912875	-0.00480756241779545\\
-44.0621471761331	0.00847324516607273\\
-43.755351018556	0.013981508059679\\
-43.4485548609789	0.0175925181690815\\
-43.1417587034018	0.0187849249050487\\
-42.8349625458247	0.0173447672149507\\
-42.5281663882474	0.013457026577619\\
-42.2213702306703	0.00760780701210706\\
-41.9145740730933	0.00058009681591642\\
-41.6077779155162	-0.00672129809939292\\
-41.3009817579391	-0.0133247924063511\\
-40.9941856003618	-0.0183844256578851\\
-40.6873894427847	-0.0212201378910564\\
-40.3805932852076	-0.0214654605054463\\
-40.0737971276305	-0.0190419278927791\\
-39.7670009700532	-0.0142284086838629\\
-39.4602048124761	-0.00755556690636894\\
-39.1534086548991	0.000196312686853162\\
-38.846612497322	0.00815849837117355\\
-38.5398163397449	0.0154089913851578\\
-38.2330201821676	0.0211431963878965\\
-37.9262240245905	0.0246938068370426\\
-37.6194278670134	0.0256665911102374\\
-37.3126317094363	0.0239093761538314\\
-37.0058355518593	0.0195890281662372\\
-36.699039394282	0.0130996982152567\\
-36.3922432367049	0.00508419634882529\\
-35.7786509215507	-0.0124036991877787\\
-35.4718547639736	-0.0202099815179935\\
-35.1650586063963	-0.0263687462992124\\
-34.8582624488192	-0.030316296091101\\
-34.5514662912421	-0.0316571449747372\\
-34.2446701336651	-0.0302588679570803\\
-33.937873976088	-0.0261897809829748\\
-33.6310778185107	-0.0197719777926935\\
-33.3242816609336	-0.0114810936000822\\
-33.0174855033565	-0.00196846529294703\\
-32.7106893457794	0.00806163262024739\\
-32.4038931882023	0.0178416386507365\\
-32.097097030625	0.0266662209086945\\
-31.790300873048	0.0338744389232559\\
-31.4835047154709	0.0389610427157123\\
-31.1767085578938	0.0415377719033003\\
-30.8699124003167	0.0414208651034755\\
-30.5631162427394	0.038566846101233\\
-30.2563200851623	0.0331352518023635\\
-29.9495239275852	0.0254023724838106\\
-29.6427277700082	0.0158051034522302\\
-29.3359316124311	0.00484130935118543\\
-29.0291354548538	-0.00689564504474305\\
-28.7223392972767	-0.0188197237891075\\
-28.4155431396996	-0.0303172160551242\\
-28.1087469821225	-0.0408441954567493\\
-27.8019508245454	-0.0498835697594942\\
-27.4951546669681	-0.0570320352910798\\
-27.188358509391	-0.0619452807606464\\
-26.881562351814	-0.0644130794671085\\
-26.5747661942369	-0.0642923316633528\\
-26.2679700366598	-0.0615719520359974\\
-25.9611738790825	-0.0562961585834856\\
-25.6543777215054	-0.0486225151944666\\
-25.3475815639283	-0.0387393438137096\\
-25.0407854063512	-0.0269206656414411\\
-24.7339892487742	-0.0134418465506201\\
-24.4271930911968	0.00136539609436426\\
-24.1203969336198	0.0171957159431884\\
-23.8136007760427	0.0337129415369049\\
-22.8932123033112	0.0845181876073084\\
-22.5864161457341	0.101012760463433\\
-22.279619988157	0.116955395841934\\
-21.97282383058	0.132175549222083\\
-21.6660276730029	0.146571084510917\\
-21.3592315154256	0.160038738817839\\
-21.0524353578485	0.172539540939042\\
-20.7456392002714	0.184027439169853\\
-20.4388430426943	0.194514308334554\\
-20.1320468851172	0.203997483358478\\
-19.8252507275399	0.212525191501662\\
-19.5184545699628	0.220123743323725\\
-19.2116584123858	0.22686372103567\\
-18.9048622548087	0.232787532177092\\
-18.5980660972316	0.237976271142855\\
-18.2912699396543	0.242478201497519\\
-17.9844737820772	0.246375999219133\\
-17.6776776245001	0.249716482633872\\
-17.370881466923	0.252577964455327\\
-17.064085309346	0.25500122891566\\
-16.7572891517686	0.257056898245708\\
-16.4504929941916	0.25877734675236\\
-16.1436968366145	0.260224204455838\\
-15.8369006790374	0.261420696184999\\
-15.5301045214603	0.262419537765183\\
-15.223308363883	0.263235165895168\\
-14.9165122063059	0.263912257871418\\
-14.3029198911518	0.264908701337141\\
-13.6893275759974	0.265562156539154\\
-12.7689391032661	0.26612806601441\\
-11.8485506305346	0.266417115675182\\
-10.0077736850719	0.266618424682996\\
-5.40583132141501	0.266848673353181\\
-4.79223900626062	0.267156959131462\\
-4.48544284868353	0.267461320923843\\
-4.17864669110645	0.267923129173859\\
-3.87185053352937	0.268635656732691\\
-3.56505437595229	0.269673953918236\\
-3.25825821837498	0.271145559784387\\
-2.95146206079789	0.273057682562396\\
-2.64466590322081	0.275234640680509\\
-2.33786974564373	0.276892761009869\\
-2.03107358806665	0.275988256292152\\
-1.72427743048934	0.267824258798541\\
-1.41748127291226	0.24335518174248\\
-1.11068511533517	0.188761757507095\\
-0.80388895775809	0.0928906474896039\\
-0.497092800181008	-0.0302501209083559\\
-0.190296642603698	-0.121858673276158\\
0.116499514973384	-0.121267078744232\\
0.423295672550466	-0.0475273439077526\\
0.730091830127549	0.0245599224622595\\
1.03688798770463	0.0558729441152366\\
1.34368414528194	0.0550645020438978\\
1.65048030285902	0.0417679510511633\\
1.95727646043611	0.0277529053633501\\
2.26407261801319	0.0170523468090948\\
2.57086877559027	0.00993716477867679\\
2.87766493316758	0.0055735255779652\\
3.18446109074466	0.00301973997778759\\
3.49125724832174	0.00159105126869008\\
3.79805340589883	0.000806194054526088\\
4.10484956347591	0.000397787095558044\\
4.7184418786303	8.13462764668316e-05\\
5.63883035136155	-3.09638964068881e-06\\
10.2407727150185	2.13621478195591e-06\\
};
\addlegendentry{$\tilde{t}=5$}

\end{axis}
\end{tikzpicture}
\end{center}

%% file: figure3.tex
%
%
\definecolor{mycolor1}{rgb}{0.00000,0.44700,0.74100}%
\definecolor{mycolor2}{rgb}{0.85000,0.32500,0.09800}%
\begin{center}
\begin{tikzpicture}

\begin{axis}[%
width=5.0in,
height=2.50in,
at={(0.758in,0.481in)},
scale only axis,
xmin=-5,
xmax=5,
ymin=-0.15,
ymax=0.3,
axis background/.style={fill=white},
legend style={legend cell align=left, align=left, draw=white!15!black,font=\footnotesize},
xlabel={$\theta$},
ylabel={$J$}
]
\addplot [color=mycolor1, mark=x, mark options={solid, mycolor1}]
  table[row sep=crcr]{%
-5.02233612444365	-0.0221750006666133\\
-4.94563708504938	-0.0177281651989878\\
-4.86893804565511	-0.0128086039662163\\
-4.79223900626062	-0.00744192713535519\\
-4.71553996686634	-0.0016565120416292\\
-4.63884092747207	0.00451693161608357\\
-4.5621418880778	0.0110457988853465\\
-4.48544284868353	0.0178960218521276\\
-4.40874380928926	0.0250324927003671\\
-4.33204476989499	0.0324194735660637\\
-4.25534573050072	0.0400209875148452\\
-4.17864669110645	0.0478011849243503\\
-4.10194765171218	0.0557246821461339\\
-4.02524861231791	0.0637568710716803\\
-3.94854957292364	0.0718641970634897\\
-3.87185053352937	0.0800144013898372\\
-3.7951514941351	0.0881767257068713\\
-3.71845245474083	0.0963220784726788\\
-3.64175341534656	0.104423162991607\\
-3.56505437595229	0.112454564666984\\
-3.48835533655802	0.120392794628904\\
-3.41165629716374	0.128216288818414\\
-3.33495725776947	0.135905362350414\\
-3.25825821837498	0.14344211673891\\
-3.18155917898071	0.150810295304837\\
-3.10486013958644	0.157995082562067\\
-3.02816110019216	0.164982844445945\\
-2.95146206079789	0.171760804451153\\
-2.87476302140362	0.178316647188471\\
-2.79806398200935	0.184638039367467\\
-2.72136494261508	0.190712058879472\\
-2.64466590322081	0.196524521859819\\
-2.56796686382654	0.202059194272874\\
-2.49126782443227	0.20729687217556\\
-2.414568785038	0.212214315961725\\
-2.33786974564373	0.216783026570802\\
-2.26117070624946	0.220967853466264\\
-2.18447166685519	0.224725427526479\\
-2.10777262746092	0.228002421689498\\
-2.03107358806665	0.230733659669548\\
-1.95437454867238	0.232840116700144\\
-1.87767550927811	0.234226887651121\\
-1.80097646988384	0.234781243399811\\
-1.72427743048934	0.234370961032939\\
-1.64757839109507	0.232843196871658\\
-1.5708793517008	0.23002427142923\\
-1.49418031230653	0.225720852010751\\
-1.41748127291226	0.219723147489137\\
-1.34078223351798	0.211810853318181\\
-1.26408319412371	0.201762670017066\\
-1.18738415472944	0.189370219283238\\
-1.11068511533517	0.174457038439046\\
-1.0339860759409	0.156902968497287\\
-0.957287036546631	0.136673577784001\\
-0.880587997152361	0.113853217533027\\
-0.80388895775809	0.0886788909263432\\
-0.72718991836382	0.0615704532791552\\
-0.650490878969549	0.0331510361046501\\
-0.573791839575279	0.00425048525855942\\
-0.497092800181008	-0.0241153334071837\\
-0.420393760786737	-0.0507946282393119\\
-0.343694721392467	-0.0745803690992979\\
-0.266995681997969	-0.0943167166548777\\
-0.190296642603698	-0.109019391733955\\
-0.113597603209428	-0.117992183326836\\
-0.0368985638151571	-0.120918936376889\\
0.0398004755791135	-0.117912342060034\\
0.116499514973384	-0.109507995361736\\
0.193198554367655	-0.0966029806828601\\
0.269897593761925	-0.0803497038001524\\
0.346596633156196	-0.0620244235559095\\
0.423295672550466	-0.0428935126085586\\
0.499994711944737	-0.0240983128558652\\
0.576693751339008	-0.00657287025043463\\
0.653392790733278	0.00899968759306091\\
0.730091830127549	0.0221942764546403\\
0.806790869521819	0.0328221628602838\\
0.88348990891609	0.0408906672079716\\
0.960188948310361	0.0465543616457094\\
1.03688798770463	0.0500659948329698\\
1.1135870270989	0.0517327484458292\\
1.19028606649317	0.051880830065608\\
1.26698510588767	0.0508293017410884\\
1.34368414528194	0.0488726051069559\\
1.42038318467621	0.0462704399696703\\
1.49708222407048	0.0432433465731732\\
1.57378126346475	0.0399723675368095\\
1.65048030285902	0.036601376928969\\
1.72717934225329	0.0332409513742791\\
1.80387838164756	0.0299729506316613\\
1.88057742104183	0.0268552339651675\\
1.95727646043611	0.023926147844314\\
2.03397549983038	0.0212085781425086\\
2.11067453922465	0.0187134716138786\\
2.18737357861892	0.0164428056111632\\
2.26407261801319	0.014392030734717\\
2.34077165740746	0.0125520363836884\\
2.41747069680173	0.0109107005319578\\
2.494169736196	0.00945408744378895\\
2.57086877559027	0.00816735407593239\\
2.64756781498454	0.00703542007085467\\
2.72426685437881	0.00604344916889499\\
2.80096589377331	0.00517718257897126\\
2.87766493316758	0.0044231579428029\\
2.95436397256185	0.00376884131392252\\
3.03106301195612	0.00320269417429131\\
3.10776205135039	0.00271419293991482\\
3.18446109074466	0.00229381461190137\\
3.26116013013893	0.00193299912867761\\
3.3378591695332	0.0016240964750871\\
3.41455820892747	0.00136030461156711\\
3.49125724832174	0.00113560271500468\\
3.56795628771602	0.000944682995472235\\
3.64465532711029	0.000782883403559254\\
3.72135436650456	0.000646122816040595\\
3.79805340589883	0.000530839736767597\\
3.8747524452931	0.000433935137540864\\
3.95145148468737	0.000352719759702858\\
4.02815052408164	0.00028486597690236\\
4.10484956347591	0.000228364163661787\\
4.18154860287018	0.0001814834074505\\
4.25824764226445	0.000142736332191973\\
4.33494668165895	0.000110847758858057\\
4.41164572105322	8.47269069801371e-05\\
4.48834476044749	6.34428336985238e-05\\
4.56504379984176	4.62028103243028e-05\\
4.64174283923603	3.2333346764446e-05\\
4.7184418786303	2.12635893710456e-05\\
4.79514091802457	1.25108357238446e-05\\
4.87183995741884	5.66792941825156e-06\\
4.94853899681311	3.92317916819707e-07\\
5.02523803620738	-3.60342357552668e-06\\
};
\addlegendentry{$\tilde{t}=0.5$}

\addplot [color=mycolor2, dashed]
  table[row sep=crcr]{%
-5.02233612444365	0.167348175293587\\
-4.86893804565511	0.177395207078199\\
-4.71553996686634	0.186796029456271\\
-4.5621418880778	0.19555008307621\\
-4.40874380928926	0.203667876608355\\
-4.25534573050072	0.21116967598745\\
-4.10194765171218	0.218084171790581\\
-3.94854957292364	0.224447078696094\\
-3.7951514941351	0.230299649771777\\
-3.64175341534656	0.235686992937921\\
-3.48835533655802	0.240656046640793\\
-3.33495725776947	0.245252974379186\\
-3.18155917898071	0.249519607164063\\
-3.02816110019216	0.25348843104272\\
-2.87476302140362	0.257175345468048\\
-2.72136494261508	0.26056919216508\\
-2.56796686382654	0.263616631991353\\
-2.414568785038	0.26620063337854\\
-2.33786974564373	0.267257683254793\\
-2.26117070624946	0.268110480652973\\
-2.18447166685519	0.268711413634834\\
-2.10777262746092	0.269001324532094\\
-2.03107358806665	0.268907624037107\\
-1.95437454867238	0.268342284926124\\
-1.87767550927811	0.267199794011495\\
-1.80097646988384	0.265355189368905\\
-1.72427743048934	0.262662378728619\\
-1.64757839109507	0.258953023805386\\
-1.5708793517008	0.254036382145975\\
-1.49418031230653	0.247700622546618\\
-1.41748127291226	0.239716267847592\\
-1.34078223351798	0.229842551466097\\
-1.26408319412371	0.217837566191396\\
-1.18738415472944	0.203473086631158\\
-1.11068511533517	0.186554796237723\\
-1.0339860759409	0.166948263190949\\
-0.957287036546631	0.144610294433444\\
-0.880587997152361	0.11962418458007\\
-0.80388895775809	0.0922358681274265\\
-0.72718991836382	0.0628862065606439\\
-0.650490878969549	0.0322329020578582\\
-0.573791839575279	0.00115434176934048\\
-0.497092800181008	-0.02927227190421\\
-0.420393760786737	-0.0578241625211859\\
-0.343694721392467	-0.0832189175394698\\
-0.266995681997969	-0.104228003566345\\
-0.190296642603698	-0.119805311658418\\
-0.113597603209428	-0.129211754307228\\
-0.0368985638151571	-0.132113843273324\\
0.0398004755791135	-0.128636245766752\\
0.116499514973384	-0.119355904437748\\
0.193198554367655	-0.105236820185627\\
0.269897593761925	-0.0875168483895887\\
0.346596633156196	-0.0675672313418287\\
0.423295672550466	-0.0467494665562347\\
0.499994711944737	-0.0262918536614611\\
0.576693751339008	-0.00720106860056546\\
0.653392790733278	0.00978497907692777\\
0.730091830127549	0.0242052130237509\\
0.806790869521819	0.0358525300708719\\
0.88348990891609	0.0447309037745205\\
0.960188948310361	0.0510032494869685\\
1.03688798770463	0.0549388747133674\\
1.1135870270989	0.0568665411550553\\
1.19028606649317	0.0571363872778266\\
1.26698510588767	0.0560917053740972\\
1.34368414528194	0.0540500219454065\\
1.42038318467621	0.0512920624857749\\
1.49708222407048	0.048056846322841\\
1.65048030285902	0.0409020297419884\\
1.80387838164756	0.0337071294532931\\
1.88057742104183	0.030305981539005\\
1.95727646043611	0.0271003553394751\\
2.03397549983038	0.0241166538058604\\
2.11067453922465	0.0213682502952244\\
2.18737357861892	0.0188586787088978\\
2.26407261801319	0.0165842640316303\\
2.34077165740746	0.0145362448812278\\
2.41747069680173	0.0127024522972787\\
2.494169736196	0.0110686119611012\\
2.64756781498454	0.00833884962404952\\
2.80096589377331	0.00622232324696181\\
2.95436397256185	0.00460182007950483\\
3.10776205135039	0.00337442718151326\\
3.26116013013893	0.00245363086850414\\
3.49125724832174	0.00149687089595574\\
3.72135436650456	0.000894192933908577\\
4.02815052408164	0.000432759107519942\\
4.48834476044749	0.000129184332559262\\
5.02523803620738	2.0499337504809e-05\\
};
\addlegendentry{$\tilde{t}=1$}

\addplot [color=black]
  table[row sep=crcr]{%
-5.02233612444365	0.262206108686623\\
-4.63884092747207	0.26394622683056\\
-3.56505437595229	0.268528916324141\\
-3.25825821837498	0.270309513151306\\
-2.95146206079789	0.272456323103476\\
-2.49126782443227	0.275842558698717\\
-2.33786974564373	0.276579150023223\\
-2.26117070624946	0.276742359269341\\
-2.18447166685519	0.276703718110903\\
-2.10777262746092	0.276400401132201\\
-2.03107358806665	0.275756144838178\\
-1.95437454867238	0.274679218796251\\
-1.87767550927811	0.273060354566192\\
-1.80097646988384	0.270770759527766\\
-1.72427743048934	0.267660411965617\\
-1.64757839109507	0.263556925840351\\
-1.5708793517008	0.258265382731972\\
-1.49418031230653	0.251569652492398\\
-1.41748127291226	0.243235861918447\\
-1.34078223351798	0.233018806403289\\
-1.26408319412371	0.220672195191021\\
-1.18738415472944	0.205963623505661\\
-1.11068511533517	0.188695010909258\\
-1.0339860759409	0.168728854945551\\
-0.957287036546631	0.146019929149396\\
-0.880587997152361	0.120650930521864\\
-0.80388895775809	0.0928690539542565\\
-0.72718991836382	0.0631186718244647\\
-0.650490878969549	0.0320635370763362\\
-0.573791839575279	0.000590725532619629\\
-0.497092800181008	-0.030211420252102\\
-0.420393760786737	-0.0591070805012412\\
-0.343694721392467	-0.0847998374917562\\
-0.266995681997969	-0.106047481140324\\
-0.190296642603698	-0.121792061015675\\
-0.113597603209428	-0.131285994502511\\
-0.0368985638151571	-0.134191902311664\\
0.0398004755791135	-0.130635920088936\\
0.116499514973384	-0.121201899093595\\
0.193198554367655	-0.106865553879162\\
0.269897593761925	-0.0888800151124638\\
0.346596633156196	-0.0686337369967207\\
0.423295672550466	-0.0475056433257146\\
0.499994711944737	-0.0267401274948407\\
0.576693751339008	-0.00735745818259659\\
0.653392790733278	0.00989405353367889\\
0.730091830127549	0.0245463373617767\\
0.806790869521819	0.0363885285092254\\
0.88348990891609	0.045423624915002\\
0.960188948310361	0.0518157603051774\\
1.03688798770463	0.0558370258308125\\
1.1135870270989	0.0578199476068813\\
1.19028606649317	0.0581189168640952\\
1.26698510588767	0.0570815846836474\\
1.34368414528194	0.0550296683561555\\
1.42038318467621	0.0522477388934428\\
1.49708222407048	0.0489782186579717\\
1.65048030285902	0.0417350085877572\\
1.80387838164756	0.034439451687776\\
1.88057742104183	0.0309871662437766\\
1.95727646043611	0.0277312250769892\\
2.03397549983038	0.0246987298656087\\
2.11067453922465	0.0219035489060628\\
2.18737357861892	0.0193495477162928\\
2.26407261801319	0.0170332548516603\\
2.34077165740746	0.014946013045968\\
2.41747069680173	0.0130756806409877\\
2.494169736196	0.0114079529095834\\
2.64756781498454	0.00861805836910623\\
2.80096589377331	0.00645083267300706\\
2.95436397256185	0.00478798953727555\\
3.10776205135039	0.00352550473317859\\
3.26116013013893	0.00257579957166953\\
3.49125724832174	0.00158518232820093\\
3.72135436650456	0.000957613173797611\\
4.02815052408164	0.000473161303301772\\
4.48834476044749	0.000149336530682831\\
5.02523803620738	2.91983688125441e-05\\
};
\addlegendentry{$\tilde{t}=2$}

\addplot [color=black, draw=none, mark=o, mark options={solid, black}]
  table[row sep=crcr]{%
-4.79223900626062	0.267159429086468\\
-4.48544284868353	0.267459390751354\\
-4.17864669110645	0.26792551027943\\
-3.87185053352937	0.268633596579652\\
-3.56505437595229	0.269676344374873\\
-3.25825821837498	0.271143440739286\\
-2.95146206079789	0.273060194455055\\
-2.64466590322081	0.275232578660757\\
-2.33786974564373	0.276895645489646\\
-2.03107358806665	0.275986541089531\\
-1.72427743048934	0.267828171518859\\
-1.41748127291226	0.243354471195272\\
-1.11068511533517	0.188767838453566\\
-0.80388895775809	0.092891187743664\\
-0.497092800181008	-0.0302431613165055\\
-0.190296642603698	-0.121860576979034\\
0.116499514973384	-0.121266518800547\\
0.423295672550466	-0.0475329816159418\\
0.730091830127549	0.0245573509309169\\
1.03688798770463	0.0558680413686714\\
1.34368414528194	0.055064135252632\\
1.65048030285902	0.0417647800190135\\
1.95727646043611	0.0277541580743348\\
2.26407261801319	0.0170498970875155\\
2.57086877559027	0.00993904745812557\\
2.87766493316758	0.00557127350600961\\
3.18446109074466	0.0030218368759849\\
3.49125724832174	0.0015888457118578\\
3.79805340589883	0.000808353681066798\\
4.10484956347591	0.000395602481148671\\
4.41164572105322	0.000183937545171098\\
4.7184418786303	7.91847364167708e-05\\
};
\addlegendentry{Analytic solution}

\end{axis}
\end{tikzpicture}
\end{center}

%% file: figure4.tex
%
%
\definecolor{mycolor1}{rgb}{0.00000,0.44700,0.74100}%
\definecolor{mycolor2}{rgb}{0.85000,0.32500,0.09800}%
\begin{tikzpicture}

\begin{axis}[%
width=5.0in,
height=2.30in,
at={(0.758in,0.481in)},
scale only axis,
xmin=-3,
xmax=3,
xtick={-3,-2,...,3},
ymin=-0.05,
ymax=0.4,
axis background/.style={fill=white},
legend style={legend cell align=left, align=left, draw=white!15!black},
xlabel={$x$},
ylabel={$\phi$}
]
\addplot [color=mycolor1]
  table[row sep=crcr]{%
-3.0068554188361	-1.21564324828682e-06\\
-1.68510810919242	0.000183001029666841\\
-1.51405845735625	0.000552775108954684\\
-1.38965871056619	0.00114169250062846\\
-1.29635890047371	0.00192196300664982\\
-1.21860905873002	0.00292285251960722\\
-1.15640918533503	0.0040485499812819\\
-1.09420931193996	0.00556903599919512\\
-1.04755940689376	0.00704444197088883\\
-1.00090950184748	0.0088781983159465\\
-0.954259596801284	0.0111462208774022\\
-0.923159660103791	0.0129433571353674\\
-0.892059723406298	0.0150042222659788\\
-0.860959786708805	0.0173634092884338\\
-0.82985985001122	0.0200593601833683\\
-0.798759913313727	0.0231341037174349\\
-0.767659976616234	0.0266330186859816\\
-0.736560039918741	0.0306047963329918\\
-0.705460103221248	0.0351012404274469\\
-0.674360166523755	0.0401773068271822\\
-0.643260229826262	0.0458910086282009\\
-0.627710261477561	0.0490058320645863\\
-0.612160293128769	0.0523029650869264\\
-0.596610324780068	0.0557902714368481\\
-0.581060356431276	0.0594755940966438\\
-0.565510388082575	0.0633667664479005\\
-0.549960419733783	0.0674714844959747\\
-0.534410451385082	0.0717973210258323\\
-0.51886048303629	0.0763515145435871\\
-0.503310514687497	0.0811409853655296\\
-0.487760546338797	0.0861722183597249\\
-0.472210577990004	0.0914511589330682\\
-0.456660609641304	0.0969830750161926\\
-0.441110641292511	0.102772478408416\\
-0.42556067294381	0.108823052850295\\
-0.410010704595018	0.115137418563439\\
-0.394460736246318	0.121717101131865\\
-0.378910767897525	0.128562347690765\\
-0.363360799548825	0.135671952732809\\
-0.347810831200032	0.143043050841937\\
-0.332260862851331	0.15067104959736\\
-0.316710894502539	0.158549280930225\\
-0.301160926153838	0.166668859175281\\
-0.285610957805046	0.175018471733014\\
-0.270060989456345	0.183584066323146\\
-0.238961052758852	0.201292488909598\\
-0.207861116061267	0.219620154664764\\
-0.161211211015074	0.24775283595518\\
-0.130111274317581	0.266488031995337\\
-0.0990113376200878	0.284766712572909\\
-0.0834613692712951	0.293597727051953\\
-0.0679114009225947	0.302143002852893\\
-0.052361432573802	0.310338583031493\\
-0.0368114642251016	0.318117941707177\\
-0.0212614958763089	0.32541251694528\\
-0.00571152752760851	0.332152308579208\\
0.00983844082118424	0.338266721230471\\
0.0253884091698846	0.343685359162698\\
0.0409383775186769	0.348338977168444\\
0.0564883458673777	0.352160613283778\\
0.07203831421617	0.355086725556303\\
0.0875882825648704	0.357058436049056\\
0.103138250913663	0.358022804983904\\
0.118688219262456	0.357934154531958\\
0.134238187611156	0.356755352785702\\
0.149788155959949	0.354459102863331\\
0.165338124308649	0.351029053685839\\
0.180888092657442	0.346460788636457\\
0.196438061006142	0.340762648309259\\
0.211988029354935	0.333956309068719\\
0.227537997703635	0.326077081128139\\
0.243087966052428	0.317173864848433\\
0.258637934401128	0.307308826126022\\
0.274187902749921	0.296556735777991\\
0.289737871098621	0.285003935051962\\
0.305287839447414	0.272747008134628\\
0.320837807796114	0.259891152818459\\
0.336387776144907	0.246548346515425\\
0.351937744493608	0.232835233474532\\
0.3830376811911	0.20477494551486\\
0.414137617888686	0.176653228774898\\
0.429687586237386	0.162847948872502\\
0.445237554586179	0.149348052809349\\
0.460787522934879	0.136243351767168\\
0.476337491283672	0.123613083832962\\
0.491887459632372	0.111524975368372\\
0.507437427981165	0.10003476898421\\
0.522987396329865	0.0891859897178957\\
0.538537364678658	0.0790101545147759\\
0.554087333027358	0.0695271812667819\\
0.569637301376151	0.0607460970869877\\
0.585187269724851	0.0526659189493821\\
0.600737238073644	0.0452768087615381\\
0.616287206422344	0.0385611724052297\\
0.631837174771137	0.0324949468271756\\
0.647387143119837	0.0270488260135151\\
0.66293711146863	0.0221895106448127\\
0.67848707981733	0.0178808071342367\\
0.694037048166123	0.014084695855594\\
0.709587016514916	0.0107622291831579\\
0.725136984863616	0.00787438689111397\\
0.740686953212409	0.00538267879271359\\
0.756236921561109	0.00324974658018196\\
0.771786889909902	0.00143972909808232\\
0.787336858258602	-8.13903015246531e-05\\
0.802886826607395	-0.0013455776240674\\
0.818436794956095	-0.00238264448718439\\
0.833986763304888	-0.00322017753229487\\
0.865086700002381	-0.00439576026846256\\
0.896186636699874	-0.00504876937732668\\
0.927286573397367	-0.00532240886016533\\
0.97393647844356	-0.00526204070909664\\
1.03613635183864	-0.00470270553125207\\
1.28493584541858	-0.00190264723195321\\
1.39378562385985	-0.00117270045688445\\
1.53373533899853	-0.00060498832569067\\
1.72033495918358	-0.000247584092626596\\
2.01578435780981	-5.81633474587839e-05\\
2.74663287020094	-2.04007753712432e-06\\
3.01098233212968	-3.93123097186532e-07\\
};
\addlegendentry{$\tilde{t}=5$}

\addplot [color=mycolor2, draw=none, mark=o, mark options={solid, mycolor2}]
  table[row sep=crcr]{%
-2.99130545048731	-8.80811946490923e-07\\
-2.91355560874362	-6.81479735309409e-07\\
-2.83580576699984	-2.16040418710151e-06\\
-2.75805592525616	-2.98032759937783e-06\\
-2.68030608351238	-2.3546217065018e-06\\
-2.6025562417686	-3.2120518587142e-06\\
-2.52480640002491	-5.29351293465652e-06\\
-2.44705655828113	-5.66887341069133e-06\\
-2.36930671653736	-5.11986465667746e-06\\
-2.29155687479367	-6.2033942662687e-06\\
-2.21380703304989	-6.66190487219609e-06\\
-2.1360571913062	-3.07503153207378e-06\\
-2.05830734956242	3.09716981616859e-06\\
-1.98055750781865	1.1958688468372e-05\\
-1.90280766607496	3.03441734144094e-05\\
-1.82505782433118	6.42257330349949e-05\\
-1.7473079825874	0.000117126238691423\\
-1.66955814084372	0.000200785317596708\\
-1.59180829909994	0.000336496085058258\\
-1.51405845735625	0.000548146883041056\\
-1.43630861561247	0.00086834964676985\\
-1.35855877386869	0.00135337734877172\\
-1.28080893212501	0.00208345211512517\\
-1.20305909038123	0.00316258083820697\\
-1.12530924863745	0.00473994273726319\\
-1.04755940689376	0.00703088187087308\\
-0.969809565149985	0.010322118611128\\
-0.892059723406298	0.014989027668423\\
-0.81430988166252	0.0215314724658224\\
-0.736560039918741	0.0305903871341386\\
-0.658810198175054	0.0429377744296437\\
-0.581060356431276	0.0594647472337422\\
-0.503310514687497	0.0811326546562658\\
-0.42556067294381	0.108817342333139\\
-0.347810831200032	0.143039701541767\\
-0.270060989456345	0.183582503893259\\
-0.192311147712567	0.228946138009339\\
-0.114561305968788	0.275710875798482\\
-0.0368114642251016	0.318117940772545\\
0.0409383775186769	0.348338975602975\\
0.118688219262456	0.357934040886127\\
0.196438061006142	0.340761836462605\\
0.274187902749921	0.296554054023785\\
0.351937744493608	0.232829518336314\\
0.429687586237386	0.162839067058791\\
0.507437427981165	0.100024157456585\\
0.585187269724851	0.0526560382340788\\
0.66293711146863	0.0221826543429611\\
0.740686953212409	0.00538008728408457\\
0.818436794956095	-0.00238093461387345\\
0.896186636699874	-0.00504356132575223\\
0.97393647844356	-0.00525447091357956\\
1.05168632018734	-0.00451175234271739\\
1.12943616193112	-0.00353493261072613\\
1.2071860036748	-0.0026265813383306\\
1.28493584541858	-0.00189436671844723\\
1.36268568716236	-0.00134337276448981\\
1.44043552890605	-0.000937339997884923\\
1.51818537064983	-0.000646096439661292\\
1.59593521239351	-0.000446122585913589\\
1.67368505413729	-0.000307820438131046\\
1.75143489588107	-0.000208363624771835\\
1.82918473762476	-0.000140933845540481\\
1.90693457936854	-9.81037678533525e-05\\
1.98468442111231	-6.71359266619653e-05\\
2.062434262856	-4.3544658319572e-05\\
2.14018410459978	-2.97860078330991e-05\\
2.21793394634347	-2.20245984121625e-05\\
2.29568378808724	-1.41544420473316e-05\\
2.37343362983102	-8.13463079740728e-06\\
2.45118347157471	-6.63818763468882e-06\\
2.52893331331849	-5.45497301196463e-06\\
2.60668315506227	-2.32035556635424e-06\\
2.68443299680595	-1.07984758734858e-06\\
2.76218283854973	-2.04589270502709e-06\\
2.83993268029342	-1.44147027603125e-06\\
2.9176825220372	3.34693680326126e-07\\
2.99543236378098	-4.94406191542396e-08\\
};
\addlegendentry{$\tilde{t}=10$}

\end{axis}
\end{tikzpicture}%

%% file: sec4c1.tex
\subsection{Comparison of numerical results for BKdV and perturbation analysis} 
\label{ssec:num_bkdv_pert}
We now consider how the numerical solution of the full nonlinear equation (\ref{eq:bkdv1}) compares with the perturbation expansion.
Considering the perturbation form given by (\ref{eq:pert_exp}), we define 
\begin{equation}
J_{\textsc{n}}(\theta,t)=\frac{\G}{\epsilon}
\left(\frac{\uv_{\textsc{n}}}{2\G^2}-\sech^2\theta\right),
\qquad
\G=\left(1+{\textstyle \frac{16}{15}}\epsilon t \right)^{-\frac{1}{2}},
\label{eq:JN}
\end{equation}
where the $\textsc{n}$ subscript denotes the numerical solution, and $\theta$ is given by (\ref{eq:thetadefn}). With $\epsilon\ll 1$, and $1\ll t\ll \epsilon^{-1}$, $J_\textsc{n}$ should agree with the asymptotic result (\ref{eq:jresult}). 
For the core region where $\theta=O(1)$, in \S\ref{ssec:num_pert}
it was seen that as $t$ increases the asymptotic solution $J$ approaches the stationary form $\JH$, with excellent agreement for $\tt>2$. 

%% file: sec4c2.tex
\begin{figure}[ht]
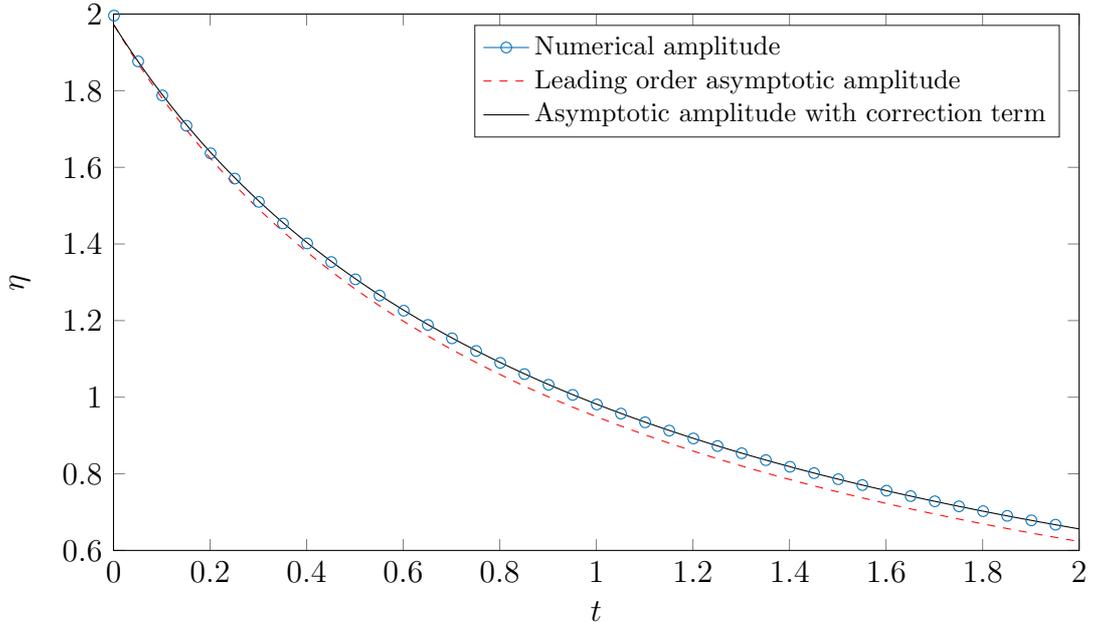

\begin{center}
\include{figure6}
\end{center}
\caption{Comparison of the maximum value of $\uv$ as a function of $\tau=\epsilon t$ obtained numerically (circles) for $\epsilon=0.1$ and the asymptotic predictions  $\uv_{\textsc{m1}}$ (solid line) and 
$\uv_{\textsc{m2}}$ (dashed line) given by (\ref{eq:umax}).}
\label{fig:maxw}
\end{figure}

The most direct check of the validity of the asymptotic predictions presented in \S\ref{sec:pert} is by considering the maximum in the waveform and its location, and comparing numerical results with the asymptotic results summarised in \S\ref{ssec:asym_summary}.
In figure \ref{fig:maxw}, the asymptotic prediction of the maximum amplitude of the disturbance is compared with the numerical results for $\epsilon=0.1$. The numerical results are compared  with the two asymptotic predictions 
\begin{equation}
\uv_{{\textsc{m1}}}=\frac{2}{1+\beta}
\left(1-
{\textstyle{\frac{2}{15}}}
\epsilon\right),
\qquad
\uv_{{\textsc{m2}}}=
\frac{2}{1+\beta}
\left(1+\left(
{\textstyle{\frac{2}{9}}}
(1+\beta)^{\frac{1}{2}}-
{\textstyle{\frac{16}{45}}}(1+\beta)^{-1}\right)\epsilon\right).
\label{eq:umax}
\end{equation}
where $\beta=1+\frac{16}{15}\epsilon t$ as previously defined.

%

As noted earlier, the term $2/(1+\beta)$ remains valid as a leading order approximation of the amplitude across the whole range of time studied, though
the first correction term becomes comparable with the leading order term when $t=O(\epsilon^{-3})$, at which point the wave amplitude is $O(\epsilon^{2})$. 
Excellent agreement is seen over a large time range as illustrated in figure \ref{fig:maxw}, and also when comparing $x_\textsc{m}$ with its numerical solution, although this is not shown here. 

%% file: figure6.tex
%
%
\definecolor{mycolor1}{rgb}{0.00000,0.44700,0.74100}%
\begin{tikzpicture}

\begin{axis}[%
width=5.0in,
height=2.8in,
at={(1.361in,0.628in)},
scale only axis,
xmin=0,
xmax=2,
ymin=0.6,
ymax=2,
axis background/.style={fill=white},
legend style={legend cell align=left, align=left, draw=white!15!black,font=\footnotesize},
xlabel={$t$},
ylabel={$\eta$}
]
\addplot [color=mycolor1, draw=none, mark=o, mark options={solid, mycolor1}]
  table[row sep=crcr]{%
0.00099999999999989	1.99606659755074\\
0.0509999999999999	1.8766547581509\\
0.101	1.78769199816484\\
0.151	1.70860569481917\\
0.201	1.63663242507451\\
0.251	1.57061650281812\\
0.301	1.50979219118742\\
0.351	1.45355231751687\\
0.401	1.40139509021309\\
0.451	1.35288697364424\\
0.501	1.30765760361567\\
0.551	1.26538439594798\\
0.601	1.22578734032334\\
0.651	1.1886180111623\\
0.701	1.15366117642953\\
0.751	1.12072343393872\\
0.801	1.0896347390554\\
0.851	1.06024355795345\\
0.901	1.03241428591298\\
0.951	1.00602501519576\\
1.001	0.98096733087781\\
1.051	0.957142129915278\\
1.101	0.934461133341094\\
1.151	0.9128430800744\\
1.201	0.892215068710351\\
1.251	0.872510254025964\\
1.301	0.853668526555233\\
1.351	0.835633657952694\\
1.401	0.818354924185737\\
1.451	0.801785838797013\\
1.501	0.785883463986345\\
1.551	0.770608111873251\\
1.601	0.755923420481345\\
1.651	0.741795699459195\\
1.701	0.728193806970455\\
1.751	0.715088856605926\\
1.801	0.702454026286192\\
1.851	0.690264423585647\\
1.901	0.678496714807278\\
1.951	0.667129759022884\\
};
\addlegendentry{Numerical amplitude}

\addplot [color=red, dashed]
  table[row sep=crcr]{%
0.00099999999999989	1.97121648398543\\
0.0190000000000001	1.93387134069835\\
0.0379999999999998	1.895949119493\\
0.0569999999999999	1.85947834451567\\
0.0760000000000001	1.82437731726376\\
0.0950000000000002	1.7905703559953\\
0.115	1.75630499627163\\
0.135	1.72331982851819\\
0.155	1.69154448375544\\
0.176	1.65941105093618\\
0.197	1.62846935644797\\
0.218	1.59865436655051\\
0.239	1.56990569337153\\
0.261	1.54087059820592\\
0.283	1.5128841925657\\
0.306	1.48468648162624\\
0.329	1.45751487214402\\
0.352	1.43131440654497\\
0.376	1.40495495584828\\
0.4	1.37954333599107\\
0.425	1.35402673844645\\
0.45	1.3294314863773\\
0.475	1.30570861045484\\
0.501	1.28191329500563\\
0.527	1.25896460695241\\
0.554	1.23598186292764\\
0.581	1.21381811910991\\
0.609	1.19165264227869\\
0.638	1.16952786637328\\
0.667	1.14820457530568\\
0.697	1.12694395872459\\
0.727	1.1064513910361\\
0.758	1.08603926257937\\
0.79	1.06573880481741\\
0.823	1.04557863174326\\
0.856	1.0261620221329\\
0.89	1.00689213024606\\
0.925	0.987792058602357\\
0.961	0.968882691978378\\
0.998	0.950182789937418\\
1.036	0.931709085539355\\
1.075	0.913476388425809\\
1.115	0.895497690672869\\
1.156	0.877784273997056\\
1.198	0.860345817086445\\
1.241	0.843190502005419\\
1.285	0.826325118786535\\
1.33	0.809755167475183\\
1.377	0.793138458119333\\
1.425	0.776852372628162\\
1.474	0.760897483389091\\
1.525	0.74496751840116\\
1.577	0.729392304681545\\
1.631	0.713887346854659\\
1.687	0.698483895202285\\
1.744	0.683467907026442\\
1.803	0.668584776301286\\
1.864	0.653858108445074\\
1.927	0.639308873092195\\
1.992	0.624955535214894\\
2	0.623233000797331\\
};
\addlegendentry{Leading order asymptotic amplitude}

\addplot [color=black]
  table[row sep=crcr]{%
0.00099999999999989	1.97133004969583\\
0.02	1.93403092872741\\
0.0390000000000001	1.89812637632198\\
0.0579999999999998	1.86353962148473\\
0.0779999999999998	1.82847792844471\\
0.0979999999999999	1.79472043459619\\
0.118	1.76219568987651\\
0.138	1.73083736586276\\
0.159	1.6990990808132\\
0.18	1.66851251476219\\
0.201	1.63901607116319\\
0.223	1.60922189605868\\
0.245	1.58050007152743\\
0.267	1.55279369630452\\
0.29	1.52485624968764\\
0.313	1.49791454296529\\
0.337	1.47080651109975\\
0.361	1.44467043705726\\
0.385	1.41945492728077\\
0.41	1.39411613581524\\
0.435	1.369674028601\\
0.461	1.34515515416266\\
0.487	1.32150653543313\\
0.514	1.29782064334887\\
0.541	1.27497664818737\\
0.569	1.25212864749057\\
0.597	1.23009280278017\\
0.626	1.20808073520726\\
0.656	1.18613152614152\\
0.686	1.16497340139518\\
0.717	1.1438964299069\\
0.748	1.12357613634668\\
0.78	1.10335156477051\\
0.813	1.08325138624842\\
0.847	1.06330181669995\\
0.881	1.04408137307334\\
0.916	1.02501578738366\\
0.952	1.0061261910939\\
0.989	0.987431650637242\\
1.027	0.968949257954177\\
1.066	0.950694225823019\\
1.106	0.932679986424886\\
1.147	0.914918291762023\\
1.189	0.897419314720878\\
1.232	0.880191749736247\\
1.276	0.863242912167958\\
1.322	0.846215916614126\\
1.369	0.829506967900925\\
1.417	0.813118192542744\\
1.467	0.796729551871143\\
1.518	0.780688263027832\\
1.571	0.764696697595856\\
1.625	0.749071767729427\\
1.681	0.733537012118123\\
1.739	0.718121013065267\\
1.798	0.703098619876391\\
1.859	0.688222365631197\\
1.922	0.673513665856922\\
1.987	0.658991465878896\\
2	0.656162895155195\\
};
\addlegendentry{Asymptotic amplitude with correction term}

\end{axis}
\end{tikzpicture}%

%% file: sec5.tex
\section{Conclusions}
\label{sec:conc}

The combination of asymptotic results and numerical results presented in \S\ref{sec:pert}
and \S\ref{sec:num} respectively provides the solution of the 
Burgers-Kortweg-de Vries (BKdV)
equation upto $t\gg O(\epsilon^{-1})$ by which time the solution is very small $O( (\epsilon t)^{-1})$. We now compare the results of the current paper with earlier analysis of equations of the general form (\ref{eqn:kdvpert}).

Asymptotic analysis of (\ref{eqn:kdvpert}) began with \cite{ott69, ott70} essentially following  the method of \S\ref{ssec:pert3}, but focussing only on the leading order term for the amplitude variation over long timescales. Applying the solvability condition at leading order yields the result
\[
\frac{\textrm{d}}{\textrm{d}t}\!\intinf \tfrac{1}{2}\uw^2 \textrm{d}x=-\epsilon\! \intinf \uw {\cal R}(\uw) \textrm{d}x.
\]
The amplitude function arises as the solution of a first order ODE in the slow time $\tau$ with its initial value fixed by the initial condition $\uu$ at $t=0$. While this analysis is not valid for $t\ll 1$, at leading order the amplitude variation is correct. In \cite{ott70} results for the amplitude variation are presented for four different dissipation processes, including the case of magnetosonic waves being damped by electron-ion collisions which is described by the BKdV equation. Their results for this case agree with (\ref{eq:maxb}) of the present analysis as $\epsilon\to 0$. The analysis of Ott \& Sudan was continued to higher order in \cite{grimshaw93}, again for the general case with dissipation 
${\cal R}(\uw)$. Here the focus is on the higher order perturbation of the wave speed  rather than the wave amplitude, though the amplitude can be readily deduced from this analysis. However when applying their general analysis to the BKdV equation an algebraic error is made. While their general result (2.20a) is valid, when this is applied to the specific case of BKdV their equation (3.8) should read
\[
\frac{\partial}{\partial T}\left( \frac{c_1}{\gamma^4}\right)=
-\frac{88\nu}{45\gamma}.
\]
With this correction, the results agree with the present treatment, which has been validated by comparison with numerical results. However the key point to note is that the analysis of  \cite{grimshaw93} does not fully determine the propagation speed as the initial value of $c_1$ remains undetermined. This has been calculated in the present paper.

We now turn our attention to the solutions of BKdV 
using inverse scattering, focussing in particular on the results of \cite{karpman79}.  
The governing equation (5.1) of \cite{karpman79} (which we denote as
(K5.1)) is identical with (\ref{eq:bkdv1}) if $u=-\uv$,  $\kappa=\G$ and 
$R(u)=u_{xx}$. The unperturbed solution is $u_s=-2\kappa^2\sech^2{z}$ and so
$ R[u_s]=-2\kappa^4(\sech^2{z})_{zz}$.
Hence (K5.9) gives the variation of wavenumber as
\[
\frac{\textrm{d}\kappa}{\textrm{d}t}=\frac{\epsilon \kappa^3}{2}
\int_{-\infty}^\infty \frac{\partial^2}{\partial z^2}\!(\!\sech^2{z})\sech^2{z} \> \textrm{d}z
=-\frac{8\epsilon \kappa^3}{15}
\]
which agrees with (\ref{eq:Gdefn}). Looking now at the amplitude of the shelf part of the tail, $w_s$,  (K5.14) and (K5.45) gives
\[
w_{Sh}=\frac{\epsilon}{4\kappa}\int_{-\infty}^\infty \frac{\partial^2}{\partial z^2}\!(\sech^2{z})\tanh^2\!{z} \> \textrm{d}z
=\frac{4\epsilon}{15\kappa},
\]
in agreement with the tail solution described in \S\ref{ssec:pert2}.
Similarly the speed of propagation given by (K5.52)
\[
\frac{\textrm{d}\xi_K}{\textrm{d}t}=4\kappa^2+\frac{\epsilon\kappa}{2}
\int_{-\infty}^\infty \frac{\partial^2}{\partial z^2}\!(\!\sech^2{z})\big{(}z\sech^2{z}+\tanh\!{z}+\tanh^2\!{z}\big{)}
\> \textrm{d}z.
=4\kappa^2+\frac{8\epsilon\kappa}{15}.
\]
agrees with the result obtained in \S\ref{ssec:pert2}.

Finally, looking  at the expression for the core,  (K5.53) should agree with the present analysis. However comparing (K5.53) with equation(3.1) of \cite{karpman78} there is an inconsistency in the coefficient of $p_1(z)$ in $J(z,z')$. After much algebraic manipulation, applying the result from \cite{karpman78} to KdVB gives the same result as our result for $J$ in (\ref{eq:jh}) with 
\[
c=\frac{1}{10}\left(1-\frac{\pi^2}{18}\right).
\]
This agrees to three significant figures with the numerical value of $c$ obtained in \S\ref{ssec:num_pert}.
Thus the results of inverse scattering theory agree exactly with the 
$1\ll t \ll \epsilon^{-1}$ asymptotic results of \S\ref{ssec:pert2}, at least for the core region, and indeed supplements the present analysis as $c$ is determined exactly. However identification of the breakdown of the  solution as time increases  from the inverse scattering analysis is unclear. In theory the exact structure of the tail should also be available from the inverse scattering analysis, though this is not discussed in \cite{karpman78,karpman79}.

In summary the present paper has produced a description of the evolution of a weakly damped soliton governed by the Burgers-Korteweg-de Vries equation, covering two different time regimes,
 $t=O(1)$ and $t=O(\epsilon^{-1})$. Comparison is made with other analyses applicable to the different time regimes and all asymptotic results have been validated by careful comparison with numerical results.
 
Of particular note is novel analysis of the region with a decaying oscillatory tail, leading to a description of the tail as a convolution of the Airy function and a characteristic function specific to the BKdV equation. The form of this function was  extracted from the numerical solution, and while other forms of perturbed KdV will have different characteristic functions describing the tail region, exactly the same methods as described here can be used to determine this function.

%% file: app_constraint.tex
\section{Integral constraints on Perturbation Equation}
\label{app:iconstraint}

In determining the solution of the linear perturbation equation, three integral constraints (\ref{eq:ivals}) are used. These constraints are derived in this appendix.
For $R(F)$ and $L(J)$ defined by (\ref{eq:LRdefn}) and $F=\sech^2\theta$, it can readily be shown that when $J$ and its derivatives tend to zero as $\theta\to\pm\infty$ then
\[
\begin{array}{rclcrcl}
\intinf R(F)\> \textrm{d}\theta&=&2\mu
&\qquad &
\intinf L(J)\> \textrm{d}\theta&=&0
\\[12pt]
\intinf FR(F)\> \textrm{d}\theta&=&
\frac{3}{2}\mu\intinf F^2\> \textrm{d}\theta-\intinf F_\theta^2\> \textrm{d}\theta
&\qquad &
\intinf FL(J)\> \textrm{d}\theta&=&0
\\[12pt]
\intinf \theta L(J)\> \textrm{d}\theta&=&\intinf (12FJ-4J) \> \textrm{d}\theta
&\qquad &
\intinf \theta R(F)\> \textrm{d}\theta&=&2\mu_1
\end{array}
\]
Integrating (\ref{eq:JfinalODE})  with respect to $\theta$ and the product of (\ref{eq:JfinalODE}) with $F(\theta)$ then gives
\[
\frac{\textrm{d}}{\textrm{d}\tt}\!\left(
\int_{-\infty}^\infty J\> \textrm{d}\theta
\right)
=
2\mu,
\qquad
\frac{\textrm{d}}{\textrm{d}\tt}\!\left(
\int_{-\infty}^\infty F J\> \textrm{d}\theta\right)
=
{\textstyle 2\left(\mu-\frac{8}{15}\right)}.
\]
Noting that $J(\theta,0)=0$ the first two integral constraints are obtained,
\[
\int_{-\infty}^\infty  J\> \textrm{d}\theta=2\mu\tt,
\qquad
\int_{-\infty}^\infty  F J\> \textrm{d}\theta=
2\left(\mu-\muno \right)\tt.
\]
Similarly, 
\begin{eqnarray*}
\frac{\textrm{d}}{\textrm{d}\tt}\!\left(
\int_{-\infty}^\infty \theta J\> \textrm{d}\theta\right)
&=&
12\int_{-\infty}^\infty FJ\> \textrm{d}\theta-
4\int_{-\infty}^\infty J\> \textrm{d}\theta-
\mu_1 \int_{-\infty}^\infty F\> \textrm{d}\theta,
\\
&=&
{\textstyle 24\left(\mu-\frac{8}{15}\right)\tt-8\mu\tt-2\mu_1.}
\end{eqnarray*}
Integrating with respect to the transformed time variable $\tt$ then gives the final integral constraint required.

%% file: app_int.tex
\section{Useful Integral Results}
\label{app:integrals}

In the main body of the paper, a number of integrals involving hyperbolic functions are used. All the integrals can be evaluated using standard techniques, their values are given together in this appendix for convenience.
Using the shorthand notation $T=\tanh\theta$ and $S=\sech^2\theta$,
\[
\arraycolsep=1.4pt
\begin{array}{rclcrclcrclcrcl}
\int_{-\infty}^\infty S^2 \>\textrm{d}\theta&=&2,
&\qquad&
\int_{-\infty}^\infty S^4 \>\textrm{d}\theta&=&
\textstyle\frac{4}{3},
&\qquad&
\int_{-\infty}^\infty S^6 \>\textrm{d}\theta&=&
\textstyle\frac{16}{15},
\\[12pt]
\int_{-\infty}^\infty \theta^3S^2T \>\textrm{d}\theta&=&
\textstyle\frac{1}{4}\pi^2,
&&
\int_{-\infty}^\infty \theta^2 S^2 \>\textrm{d}\theta&=&
\textstyle\frac{1}{6}\pi^2,
&&
\int_{-\infty}^\infty \theta S^2T \>\textrm{d}\theta &=&1,
\\[12pt]
\int_{-\infty}^\infty \theta S^4 T\> \textrm{d}\theta&=&\frac{1}{3},
&&
\int_{-\theta_\textsc{m}}^\infty (1-T) \>\textrm{d}\theta&=&
2\theta_\textsc{m},
&&
\int_{-\theta_\textsc{m}}^\infty \theta (1-T) \>\textrm{d}\theta&=&
\textstyle\frac{1}{12}\pi^2-{\theta_\textsc{m}}^2.
\end{array}
\]

%% file: app_step.tex
\section{Convolution Integrals involving Airy functions}
\label{app:tophat}

Defining the convolution $R=f*\Ai$ and
\[
J(z)=\int_{-\infty}^z R(z') \> dz',
\qquad
I_1(z)=\int_{-\infty}^z J(z') \> \textrm{d}z',
\qquad
I_2(z)=\int_{-\infty}^z z'J(z') \> \textrm{d}z',
\]
it is postulated
that as long as $f(x)$ decays sufficiently rapidly as $\vert x\vert\to\infty$ so that 
\[C_0=\intinf f(x)\> \textrm{d}x,\quad 
C_1=\intinf xf(x)\> \textrm{d}x,\quad 
C_2=\intinf x^2 f(x)\> \textrm{d}x,\]
then
\begin{equation}
J\sim C_0, \qquad
I_1(z) \sim C_0 z- C_1,
\qquad
I_2(z) \sim
{\textstyle \frac{1}{2}} \left(C_0 z^2- C_2\right),
\label{eq:jblim1}
\end{equation}
as $z\to\infty$.
This corresponds to (\ref{eq:jhlim}) with appropriate change of variables.
Here it is verified that the result is true 
when $f(x)=1$ for $a<x<b$ and zero elsewhere. Since all the results are linear in $f(x)$, this then validates the result for all  piecewise constant functions with compact support.

We begin by defining the function $I_A=\int_{-\infty}^\infty \Ai(z')\> \textrm{d}z'$ from which it follows that 
\begin{eqnarray*}
q(z)&=&\int_{-\infty}^z I_A(z') \textrm{d}z'
=zI_A-\Ai'
\\
r(z)&=&\int_{-\infty}^z q(z') \textrm{d}z'
=\thalf (z^2I_A-z\Ai'-\Ai)
\\
s(z)&=&\int_{-\infty}^z z' q(z') \textrm{d}z'
={\textstyle\frac{1}{3}}
(z^3I_A-z^2\Ai'-z\Ai+I_A).
\end{eqnarray*}
Taking the limit as $z\to\infty$, and using the asymptotic forms
\cite{abramowitz64} for Airy functions gives
$I_A\sim 1$,
$q\sim z$,
$r\sim \thalf z^2$
and
$s\sim {\textstyle\frac{1}{3}}(z^3+1)$.
Hence
\[
R(z)=
\int_a^b \Ai(z-y) dy=I_A(z-a)-I_A(z-b)
\qquad
\mbox{and so}
\qquad
J(z)=q(z-a)-q(z-b).
\]
Thus $J\sim b-a$ and similarly, after some algebra,
\[
I_1\sim
(b-a)z-\thalf (b^2-a^2),
\qquad
I_2\sim
\thalf (b-a)z^2-{\textstyle\frac{1}{6}}(b^3-a^3).
\]
Finally, substituting for $f(x)$ in the expressions for $C_i$,
\[
C_0=b-a
\qquad
C_1=\thalf (b^2-a^2)
\qquad
C_2={\textstyle\frac{1}{3}} (b^3-a^3),
\]
and the results (\ref{eq:jblim1}) are validated.